# An Experimental Investigation of Preference Misrepresentation in the Residency Match

Alex Rees-Jones and Samuel Skowronek[*]

**Abstract:** The development and deployment of matching procedures that incentivize truthful preference reporting is considered one of the major successes of market design research. In this study, we test the degree to which these procedures succeed in eliminating preference misrepresentation. We administered an online experiment to 1,714 medical students immediately after their participation in the medical residency match—a leading field application of strategy-proof market design. When placed in an analogous, incentivized matching task, we find that 23% of participants misrepresent their preferences. We explore the factors that predict preference misrepresentation, including cognitive ability, strategic positioning, overconfidence, expectations, advice, and trust. We discuss the implications of this behavior for the design of allocation mechanisms and the social welfare in markets that use them.

People often have strong incentives to lie about their preferences. These incentives are unfortunate, since market organizers must commonly make decisions based on the preferences that individuals report. Auction prices are often determined based on bids, but potential buyers may not bid their true valuation. Employees are often hired based on interviews, but job seekers may feign interest for the positions available. Students are often assigned to schools based on reported school preferences, but applicants may be incentivized to list an attainable school as their favorite. In environments like these, economists have devoted substantial effort to mitigating this problem by designing *strategy-proof mechanisms* that render truthful preference reporting incentive compatible. With such a mechanism in place, market participants who understand how outcomes are determined will see that there is no benefit to lying.

---

[*] Rees-Jones: Operations, Information, and Decisions Department, The Wharton School and NBER (email: alre@wharton.upenn.edu). Skowronek: Operations, Information, and Decisions Department, The Wharton School (email: samsko@wharton.upenn.edu).

A growing body of evidence suggests that individuals misrepresent their preferences in incentive-compatible environments despite the futility of such efforts. Imperfect truth-telling has been documented in laboratory experiments studying sealed-bid and clock auctions (1), in willingness-to-pay elicitations (2), and in applications of school-choice matching mechanisms (3–7). This work has informed recent theoretical advances aimed at characterizing mechanisms that are "obviously strategy-proof" to relatively unsophisticated decision-makers (8). In many contexts, attendance to this criteria yields comparatively easy-to-understand mechanisms; however, in the context of stable two-sided matching mechanisms, no obviously strategy-proof options exist (9). An immediate implication is that, in matching environments where stability is required, we must rely on a degree of sophistication in market participants for optimal behavior to emerge.

Particularly in the context of student matching markets, these findings can be viewed as troubling. A key argument motivating the adoption of strategy-proof school-choice mechanisms is that they "level the playing field" (10). In algorithms with a nontruthful optimal strategy, strategically savvy—and disproportionately affluent—students are given an undue advantage at the expense of students who report their preferred schools truthfully. If strategy-proof mechanisms result in all participants reporting truthfully, this undesirable outcome is averted. However, if the inability to understand optimal strategies extends to cases where the optimal strategy requires no "gaming" of the system, an unleveled playing field remains. Understanding the prevalence and correlates of such mistakes then becomes crucial for assessing the fairness, and indeed the broader welfare consequences, of the allocations that these mechanisms generate.

Unfortunately, directly assessing the prevalence and correlates of preference misrepresentation is fundamentally challenging. In the field settings where these mechanisms are adopted, preferences are unobservable. Absent observing true preferences, the veracity of reported preferences cannot be directly assessed.[†] Experimenters have sidestepped this difficulty in the laboratory by using simplified matching scenarios to assign preferences. However, by restricting empirical investigations to the lab, such work can only document suboptimal behavior in unfamiliar and minimally incentivized tasks completed by populations different from the ones facing these mechanisms in the field. On the one hand, these external validity concerns potentially mitigate the worry that the observed failure of optimal reporting extends to the policy applications of primary concern. On the other hand, if misrepresentation persists in populations whose lives are affected by their performance in these mechanisms, the design and deployment of these mechanisms may require considerable revision.

In this study, we aim to achieve the benefits of the lab-experimental approach to detecting failures of truth-telling while simultaneously studying the behavior of a highly incentivized and highly trained population of direct policy relevance. We deploy a large-scale online experiment to 1,714 medical students participating in the 2017 National Resident Matching Program (NRMP), a system in which graduating medical students submit their preferences over residency programs to be used to determine their placements. The NRMP utilizes a modified version of the deferred acceptance algorithm (15, 16), a matching mechanism that is strategy-proof for students and is increasingly adopted for school assignment (17). The NRMP constitutes a flagship application of matching theory, and remains one of the most carefully designed and extensively studied two-sided matching markets in existence.‡ Our online experiment puts NRMP participants through a simple incentivized matching task in which truth-telling can be easily assessed. By deploying this study immediately after the NRMP match, and by transparently applying the same mechanism used by the NRMP, we are able to directly study the prevalence and correlates of preference misrepresentation in the precise population of interest.

---

[†] Despite this difficulty, some attempts to assess rates of truth-telling in field settings have been made. To sidestep the difficulty of observing true preferences, researchers have relied on either unincentivized survey reports of self-assessed truthful behavior (11) or have examined specific types of reported preferences that are so anomalous that they cannot be plausibly explained by preference heterogeneity (12–14).

We document widespread failure to pursue the incentivized strategy of truth-telling. Over 23% of experimental participants misrepresent preference in our matching task, despite using this mechanism to make a career-altering decision mere days before.

We additionally examine the predictors of misrepresentation, shedding light on both the factors that contribute to this behavior and the features of individuals who bear the costs. The tendency for suboptimal behavior is associated with both the strength of the students' strategic position (measured by randomly-assigned test scores in the matching task) and by the students' cognitive reasoning abilities (measured by Raven's Matrices deployed after the matching task). Beyond metrics associated with student quality, the tendency for suboptimal behavior is associated with students' overconfidence, with the pursuit and availability of advice in the lead up to the NRMP match, and with students' trust in residency programs to rank students according to quality. These results identify the individuals who gain and lose from the complexity of the existing system, give guidance on the best practices for training market participants to engage with complex mechanisms, and critically inform the study and design of matching markets. We further discuss these implications in section IV.

---

‡ Independent of its relation to the mechanism design literature, the NRMP is of intrinsic importance to a large labor market. In 2017 alone, the NRMP received 35,696 preference lists from applicants vying for 31,757 positions (18).

**I. Study Population and Sample Recruitment**

We solicited participation in our study by recruiting medical schools to present our recruitment materials to their NRMP participants, following recruitment protocol derived from previous survey investigations of medical students (19). To do so, we contacted representatives of all 147 medical schools accredited by the Association of American Medical Colleges (AAMC) located in the United States and Puerto Rico. As a result of our initial outreach and subsequent follow-up, we were able to successfully recruit 25 medical schools (see SI Appendix table S1 and fig. S1). These 25 schools vary widely in class size (min=41, max=328), location, and competitiveness. Compared to the full population of accredited medical schools, we find no statistically significant differences between participating and non-participating schools on total enrollment, average MCAT performance, average undergraduate GPA, acceptance rates, US News and World Report Research Rankings, or gender composition (see SI Appendix table S2).

Shortly after the deadline for submission of residency preferences to the NRMP, participating schools forwarded our recruitment email to their graduating student body. This email asked students to participate in an anonymous 10-minute survey about decision-making in the NRMP match. Students were further told that they would earn an Amazon.com gift card valued between $5 and $50 with an expected value of $21 for participating in the survey. All data was collected prior to the NRMP's announcement of the results of the match.

Approximately 3,300 graduating medical students (17.1% of all graduating medical students from AAMC accredited schools) received an email with our survey link. Participant demographics are summarized in SI Appendix table S3. Our analysis is based on the 1,714 students (approximately 51.9% of the students contacted) who both completed the survey and passed all exclusion criteria (see SI Appendix table S4 and fig. S2).

**II. Experimental Design**

All experimental materials are presented in the SI Appendix; we summarize the key measures below. Our materials were reviewed by the University of Pennsylvania Institutional Review Board (IRB) and ruled exempt from IRB review (as authorized by 45 CFR 46.101(b), category 2). Informed consent was elicited on the first page of the web survey.

**II.A Incentivized Matching Task**

Participating students were presented with an incentivized matching task. The prompt for this task explained: "In this exercise, you will go through a matching process much like the NRMP match. You will attempt to match to one of five hypothetical residency programs, and the payment you receive for taking this survey will dependent on where you match. We will apply the standard algorithm that was used by the NRMP;

as a reminder, an example of how this algorithm works is available here." The underlined term hyperlinked to NRMP training materials. Since students receive significant training and advice regarding this algorithm in the lead-up to participating in the NRMP match, we did not elaborate further on the functioning of this mechanism.

In each simulation, 50 students applied to 5 residency programs, each with 10 positions available. The preferences of both the programs and the other students are simulated according to guidelines communicated to the participant. We explained that all students agree on the same ranking of residency programs. We also explained that residency programs based their preferences on several factors, with students' "Hypothetical Standardized Test" (HST) scores being an important one. Based on the manner in which programs' preferences were simulated, every student had some possibility of matching to every program. This renders nontruthful preference reporting a strictly suboptimal strategy for maximizing expected payoff.

In order to communicate the desirability of different residency programs, participants were presented with a table (fig. 1). For each program, this table reported both the average HST score of the admitted students and the value of the Amazon.com gift card that participants would receive if they matched. Participants were also told that they would earn $5.00 if they did not match to any program. The payment received from this matching process was the sole compensation provided for participation.

After this explanation of the matching task, students submitted their rank-order list (ROL) using a series of dropdown menus. Participants were told that they must apply to at least one program but could forego latter applications if they wished.

We will refer to ROLs that list all five residencies in order of their compensation as *optimal* or *truthful*, and those that do not as *suboptimal* or *misrepresented.* This labeling relies on the assumption that participants prefer more money to less. While that assumption is both standard and reasonable, under some conditions it could fail. For example, failure could arise if subjects prefer to earn less money because they value leaving money to the experimenter or to the simulated students that they compete against. We consider this possibility unlikely. Failure could also arise if subjects value not only the monetary payoffs but also anticipation or disappointment. We further discuss this latter possibility in our tests of possible correlates below. While it is necessary to rule out non-standard preferences to ensure that misrepresented ROLs identify confusion about incentives, misrepresentation stemming from either source would be viewed as anomalous from the perspective of standard matching theory.

**II.B Correlates of Suboptimal Reporting**

We preregistered our interest in five groups of correlates of suboptimal reporting, all proposed and discussed in prior literature (for a summary, see (20)). Not all of the variables that we examine are experimentally manipulated, and consequently not all analyses can be interpreted as estimating causal relationships. However, some of the associations help distinguish between potential factors driving the suboptimal behavior of interest. Furthermore, different predictors of misrepresentation suggest different welfare costs of this behavior, and the necessary approaches to reduce it. We motivate each factor of interest below, and explain its measurement in the context of our study.

*Student quality:* The welfare consequences of misrepresentation can be significantly influenced by its correlation with student quality (21). Two distinct channels, conflated in the field but separable in our experiment, may generate such a correlation. First, students with comparatively low grades or test scores are often placed at a strategic disadvantage for obtaining a desirable match. This might result in attempts to misrepresent preferences as a means to compensate, or might lead students to fail to list desirable programs under the belief that they are unobtainable. Second, students in this position might also have comparatively low cognitive ability, which increases the probability of incorrectly identifying the optimal strategy in lab experiments utilizing this algorithm (22). Our experiment contains measures that allow us to study each channel separately.

To examine the impact of strategic positioning, participants were randomly assigned an HST percentile score. This score influenced each participants' ranking in residency program preferences, and thus their strategic position.

To examine the impact of cognitive ability, we presented participants with a test of spatial reasoning. We gave participants five minutes to complete seven Advanced Raven's Progressive Matrices (23), a test widely used to assess logical reasoning ability (24). Of course, medical students with low cognitive ability relative to their peers likely have substantially higher average cognitive ability than typical populations facing matching mechanisms (e.g., school children and their parents). Care is warranted when extrapolating our results onto other such populations.

*Overconfidence:* Overconfidence is a prevalent trait among physicians (25), and is commonly thought to broadly generate decision errors (26). Furthermore, recent research demonstrates that this bias affects suboptimal reporting in the related, but gameable, Boston mechanism (27). We generate a measure of overconfidence in the course of conducting our test of logical reasoning ability. After completing the Raven's Matrices, participants were asked to think about other medical students participating in this survey and to estimate the percentage of participants that they outperformed (slider scale: min: *0%,* max: *100%)*. We code participants as overconfident if their forecast of their performance exceeds their actual percentile rank—in the language of Moore and Healy, this is a measure of overplacement (28). A secondary, but similar, measure of overconfidence is available from students' report and assessment of their MCAT performance. Participants were asked to report their MCAT score and then estimate the percentage of other MCAT takers who received a lower score than they received in the year that they took the MCAT (slider scale: min:*0%,* max:*100%*).

*Desire to rank the expected outcome highly:* If students derive utility from the anticipation of matching to a program that they rank highly, or if they expect to experience disappointment from matching to a program that they did not rank highly, then students may be motivated to submit nontruthful preference orderings that manage these anticipations. In this case, misrepresentation need not be irrational: in the presence of such belief-based utility functions, the deferred acceptance algorithm is not strategy-proof.

We test for the influence of expectations on misrepresentation by randomly varying the salience of the participants' expected match before they submit their ROLs. Before proceeding to the submission page, we randomly assigned half of participants to indicate the residency where they expected to match. We reminded them of their expected match in the list submission prompt.

*Pursuit and availability of advice:* When mechanisms are sufficiently difficult to understand, participants may be significantly influenced by advice (or their tendency to seek it) (29, 30). To examine the role of advice, we requested that participants check all of the sources that provided them with advice regarding their NRMP submission from the following list: 1) Current and/or past medical students who participated in the NRMP, 2) Participant's medical school, 3) The NRMP website, and 4) Other sources. Participants then specified the advice they received from each entity in a free-response text box and rank ordered them based on the level of influence each had on their NRMP submission.

*Mistrust of other market participants:* In many mechanisms, a particular action (such as truth-telling) may be an optimal strategy if and only if all other market participants similarly pursue optimal play. Note that this is not the case in the deferred acceptance algorithm that underlies the NRMP matching algorithm: truth-telling is optimal regardless of the action of other market participants (31, 32). However, if participants misunderstand this distinction, or if they harbor mistrust of other market participants that leads them to doubt the credibility of the matching agency, suboptimal behavior could arise (33).

We asked participants whether they trusted the players in the NRMP matching market. Participants indicated 1) whether they trusted that the residencies that they rank ordered in their NRMP submission would rank order medical students based on a truthful assessment of their quality, 2) whether they trusted other medical students to submit a truthful rank ordering of their preferences to the NRMP and 3) whether they trusted the NRMP to run the matching algorithm honestly (all questions, 1=*Yes*, 0=*No*).

**III. Results**

We examine the data in three stages. First, we catalogue the various ways participants submitted their ROL of the residency programs in the simulated match and document the monetary consequences of suboptimal behavior. Second, we provide evidence that behavior in our experiment is associated with known proxies for misunderstanding in the NRMP match. Third, we examine the correlates of suboptimal behavior.

**III.A Documenting Suboptimal Behavior**

To apply optimally in the incentivized matching exercise, participants must rank residencies in order of their monetary value. Applications are suboptimal if participants *shorten* their ROL by not ranking all pertinent programs or *permute* their ROL by ranking their listed programs in an order that does not reflect the monetary payoffs.

We find that 23.3% (N=399) of participants applied suboptimally. As shown in Figure 2, 64.7% of participants who submitted a suboptimal ROL permuted the list of residency programs (15.1% of total N) but applied to all five programs, 28.3% (6.6% of total N) of participants shortened their ROL, and 7.0% (1.6% of total N) both shortened and permuted their ROL. (See SI Appendix figs. S3 and S4 for analysis of the programs removed and misordered in suboptimal ROLs.)

Failure to submit the optimal ROL was costly. Participants who submitted a suboptimal ROL earned $18.20 on average, 21.2% less than the average earnings of participants who submitted an optimal ROL, $22.80 (t=-5.43, *p*<.001).[§] However, this difference cannot be entirely attributed to the effect of misrepresentation because participants' assigned HST scores affect both their earnings and their propensity to misrepresent preferences. As we document in section III.C, misrepresentation becomes less common among students assigned comparatively high HST scores. The rate of misrepresentation varies from 28.6% in the second lowest decile up to 14.0% in the second highest decile. The solid lines in Figure 3 show that the average difference in experimental earnings between optimal and suboptimal participants is most dramatic for those assigned a comparatively high HST score, but persisted across the distribution of assigned strategic positions (for statistical tests, see SI Appendix table S5). While all students in the experiment are incentivized to truthfully report preferences, these results illustrate that the strength of incentives varies based on the student's position in the market. This variation in incentives is a key feature of this class of matching problems, and a possible channel driving the hypothesized association between misrepresentation and student ability. A desirable student has a strictly larger set of possible match partners, which results in larger differences between the best and the worst outcomes that are possible from different reporting strategies.

---

[§] SI Appendix fig. S5 assesses the rate of costly misrepresentation at the individual level.

### III.B Validation of Experimental Behavior

We conduct three validation exercises to confirm that behavior in our experiment proxies for misunderstanding of incentives in the real residency match.

First, we test for differences in the rate of misrepresentation in our experiment as a function of self-reported truth-telling status in the NRMP. We find that students who report misrepresenting preferences to the NRMP are 9.4 percentage points more likely to misrepresent preferences in our experiment (22.6% vs. 33.0%, $\chi^2$=6.19, $p$=.013).

Second, we test the correlation between misrepresentation in our experiment and the propensity for students to submit comparatively short preference lists to the NRMP. Short lists are a known proxy for suboptimal preference reporting and are actively discouraged in NRMP training materials (34). We regressed participants' likelihood to shorten their experimental ROL (1=shortened, 0=not) or to permute their experimental ROL (1=permuted, 0=not) on the number of programs participants ranked in their NRMP submission. We find that participants who submitted either shortened or permuted ROLs submitted shorter ROLs to the NRMP (shortened: B= -0.78, SE=0.120, p<.001; permuted: B=-0.19, SE=0.089, p=.038) (see SI Appendix fig. S6 for details).

Third, we examine differences in truth-telling rates across students who do, and do not, expect to match to their top-ranked program in the NRMP match. We find that participants who expected to match to their top NRMP match choice (N=1,157; 67.5% of sample) were significantly less likely to submit an optimal ROL in the incentivized exercise (75.1%) compared to participants who did not hold this expectation (80.1 %) ($\chi^2$=5.19, $p$=.023). This result is consistent with our measure capturing a belief that optimal strategies involve strategically ranking attainable schools highly, a key component of optimal strategies in related, but manipulable, mechanisms (e.g., the Boston Mechanism).

In summary, our experimental measure validates well with proxies for suboptimal preference reporting in the field.

**III.C Examining the Correlates of Suboptimal Behavior**

Figure 4 summarizes the full battery of tests of the correlates of suboptimal preference reporting. Plotted are estimated average marginal effects (AMEs) derived from a logit model predicting the outcome of submitting a truthful preference ordering. Panel B presents the estimate for each univariate model, predicting truth-telling with only the single variable represented in that row.¶ These results provide guidance on the features of students who do, or do not, face difficulties in pursing the optimal strategy. Panel A presents estimates obtained from the complete model, including the entire battery of predictors. These provide clearer guidance of the role of each considered correlate, holding all else equal. We normalize all continuous variables in this analysis, so their coefficients may be interpreted as the association of a one-standard-deviation increase in the relevant variable.

*Student quality:* Prior work examining unincentivized assessments of truth-telling status (11) or a subclass of egregious mistakes (12, 13) has provided evidence that students with better grades are less likely to misrepresent their preferences. We replicate this finding with our incentivized experimental measure. Participants with higher MCAT scores were significantly more likely to submit an optimal ROL: a one-standard deviation increase in MCAT scores is associated with a five percentage point increase in the rate of truth-telling (AME=0.05, SE=0.010, *p*<.001).

---

¶ SI Appendix fig. S7 reports these analyses using the self-reported measure of truth-telling. In accordance with our preregistration plan, we treat these results as secondary.

In field settings, an association with test scores can be jointly influenced by both response to a poor strategic position and by differences in logical reasoning ability. These channels are separable in our experiment, and we find evidence that both channels are active. Participants assigned to higher HST scores were more likely to submit an optimal ROL (AME=0.04, SE=0.010, *p*<.001). Furthermore, participants who performed better on the Raven's Matrices task were more likely to submit an optimal ROL (AME=0.03, SE=0.010, *p*=.002). As indicated in Figure 4A, these estimates maintain comparable magnitudes and statistical significance while controlling for the full battery of correlates. In summary, the characteristics of high-performing students are useful individual predictors of truth-telling behavior, even when holding other factors constant.

*Overconfidence:* Examined in isolation, participants exhibiting overconfidence on the Ravens' task were two percentage points more likely to submit the optimal preference ordering, although we cannot reject the null hypothesis of no effect (difference of proportion *z*=-0.91, *p*=.365). This difference becomes greater in both magnitude and statistical significance in the complete model, at least partially due to eliminating the offsetting effect of our overconfidence measure's strong negative association with Raven's task performance (r = -0.59, *p*<.001). All else equal, overconfident participants were more likely to submit optimal ROLs compared to non-overconfident participants (AME=0.08, SE=0.028, *p*=.005). Similarly, controlling for MCAT performance, participants who overestimate the percentile of their MCAT score submit an optimal ROL at a significantly higher rate (AME=0.09, SE=0.027, *p*=.001).[**]

*Desire to rank the expected outcome highly:* We find no support for an effect of our expectations-salience manipulation. No significant difference is found in the propensity to report truthfully as a function of the expectations condition (difference of proportion *z*=0.33, *p*=.743).

*Pursuit and availability of advice:* Medical students usually seek out and receive advice from many sources about how to maximize their chances for admission to a top residency program. Consistent with this tendency, 71.6% of participants report receiving advice from their medical school, 62.3% reported receiving advice from other students, 40.6% reported receiving advice from the NRMP website, and 23.6% reported receiving advice from other sources. We find that the pursuit and receipt of advice is significantly associated with the likelihood to submit an optimal ROL. Participants showed an increased likelihood to submit an optimal ROL when they reported receiving advice from their medical school (AME=0.08, SE=0.024, *p*=.001), other students (AME=0.04, SE=0.021, *p*=.043), the NRMP website (AME=0.12, SE=0.020, *p*<.001), or other sources (AME=0.09, SE=0.022, *p*<.001). As shown in Figure 4A, the estimates associated with receiving advice from the NRMP website (AME=0.10, SE=0.021, *p*<.001) and other sources (AME=0.07, SE=0.023, *p*=.002) remain largely unchanged in the complete model while those associated with receiving advice from other students (AME=-0.01, SE=0.023, *p*=.565) and from participants' medical school (AME=0.036, SE=0.025, *p*=.158) attenuate. Similar results are found when regressing truthtelling status on all advice sources simultaneously, excluding all other factors (SI Appendix table S6). We present extensive exploratory text analysis of the reports of advice received, and show the effects of source influence, in the SI Appendix (table S7, figs. S8-S12).

---

[**] Since overconfidence is typically associated with decision errors, the positive correlation documented here may be viewed as surprising. However, this positive relationship could naturally arise from our results on student quality, assuming that overconfidence leads students to overestimate the strength of their strategic position.

*Mistrust of other market participants:* While 97.3% of participants trusted the NRMP to run the algorithm honestly, 63.4% of participants did not trust other students to submit a truthful ROL and 42.0% of participants did not trust their residencies to rank order students fairly. We find that participants' likelihood to submit an optimal ROL decreased by five percentage points if they trusted residencies to rank order graduating medical students based on an honest assessment of their quality (AME=-0.05, SE=0.020, *p*=.017), but that neither trust in other students (AME=0.00, SE=0.021, *p*=.982) nor in the NRMP significantly affected performance (AME=0.05, SE=0.067, *p*=.446). These effects remain largely unchanged in the complete model, or when regressing truthtelling status on all trust measures simultaneously, excluding all other factors (SI Appendix table S8).

**IV Discussion**

A large literature in economics has focused on the design of mechanisms that incentivize truth-telling, and a large theoretical literature has assumed that behavior in these mechanisms is ultimately truthful. In this paper, we have demonstrated that highly trained and incentivized participants in a flagship application of mechanism design appear to misunderstand these incentives at a substantial rate. Furthermore, this behavior is critically tied to student quality, to overconfidence, to the pursuit and the sources of available advice, and to trust in residency programs.

An immediate implication of our results is that there is room for training programs to help medical students avoid harming themselves through attempts to game the system. As we document, students receiving advice from credible advisors are significantly more likely to behave optimally. At the same time, students reliant on the advice from other students—a potentially non-credible source—are no better, and potentially worse, at finding the optimal strategy. These results converge with evidence from the lab suggesting that trust in the "folk wisdom" of other market participants may be misplaced (35). Indeed, as we document in the SI Appendix (figs. S8 and S9), free-response descriptions of the advice provided from all sources reveals that a substantial fraction of recommended strategies are misguided. Attempts to better direct students to credible, high-quality advice are clearly needed.

Because different groups face different rates of misrepresentation, and because misrepresentation harms the outcomes of those who pursue it, the use of this mechanism will ultimately favor the groups who best understand it. To the extent that misunderstanding is driven by student ability, this can be desirable. Prior research highlights the potential for misunderstanding of the deferred acceptance algorithm to serve as a screening device and facilitate matching the best students to the best schools (21).[††] However, our results suggest that factors beyond ability are favored through this channel. Overconfident students, students receiving credible advice, and students distrustful of residency programs are the net beneficiaries in our experiment—an outcome that is likely undesirable when compared to the outcome that would arise under universal truth-telling. Similar results can arise over more basic demographics: for example, in our data, women are eight percentage points more likely to misrepresent their preferences ($\chi^2$=16.85, $p$<.001), implying that men are the net beneficiaries of the presence of misrepresentation in this market. For reasons of both fairness and market efficiency, utilization of a mechanism that systematically rewards groups for factors independent of ability is typically viewed as undesirable. Further interventions to mitigate these effects are likely worthwhile, but to the extent that some residual misunderstanding is unavoidable, we encourage further research aimed at formally assessing the comparative performance of different matching mechanisms in the presence of persistent misunderstanding.

---

[††] While screening on ability is the most natural consideration in the NRMP, screening on other dimensions can become important in other markets. For example, a recent study of the Hungarian college matching system finds that relatively affluent students are more likely to report preferences that suboptimally forego chances for scholarships, ultimately resulting in better targeting of financial aid to those in need (13).

**Acknowledgments:** We thank Katy Milkman, Maurice Schweitzer, and Ran Shorrer for helpful comments. We thank Melissa Beswick for excellent research assistance, and Vincent Conley for assistance in coding our experiment. We thank the Wharton Behavioral Lab and the Wharton Risk Center for financial assistance.

**Fig. 1. Residency Information for Simulated Residency Match**.

| Rank | Residency program | Average HST percentile score of admitted students | Amazon.com gift card value |
|---|---|---|---|
| 1. | Maplecrest | 80$^{th}$ percentile | $50.00 |
| 2. | Birch Hill | 65$^{th}$ percentile | $25.00 |
| 3. | Elm South | 50$^{th}$ percentile | $15.00 |
| 4. | Hickory Bridge | 35$^{th}$ percentile | $10.00 |
| 5. | Pine Peak | 20$^{th}$ percentile | $7.50 |

**Notes:** This table was displayed to participants to communicate the desirability of different programs. The desirability was communicated in two ways: first, by the average scores on the "Hypothetical Standardized Test" (HST) of students admitted to each residency; and second, by the value of the gift card that participants would earn by matching to that program.

**Fig. 2. Classification of Truth-telling Behavior.**

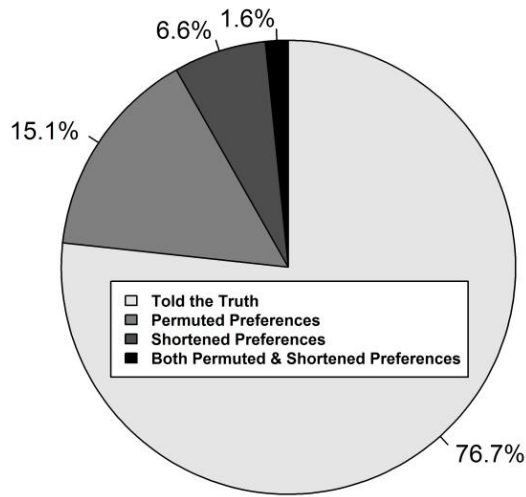

**Fig. 3. Monetary Losses Associated with Suboptimal Preference Reporting.**

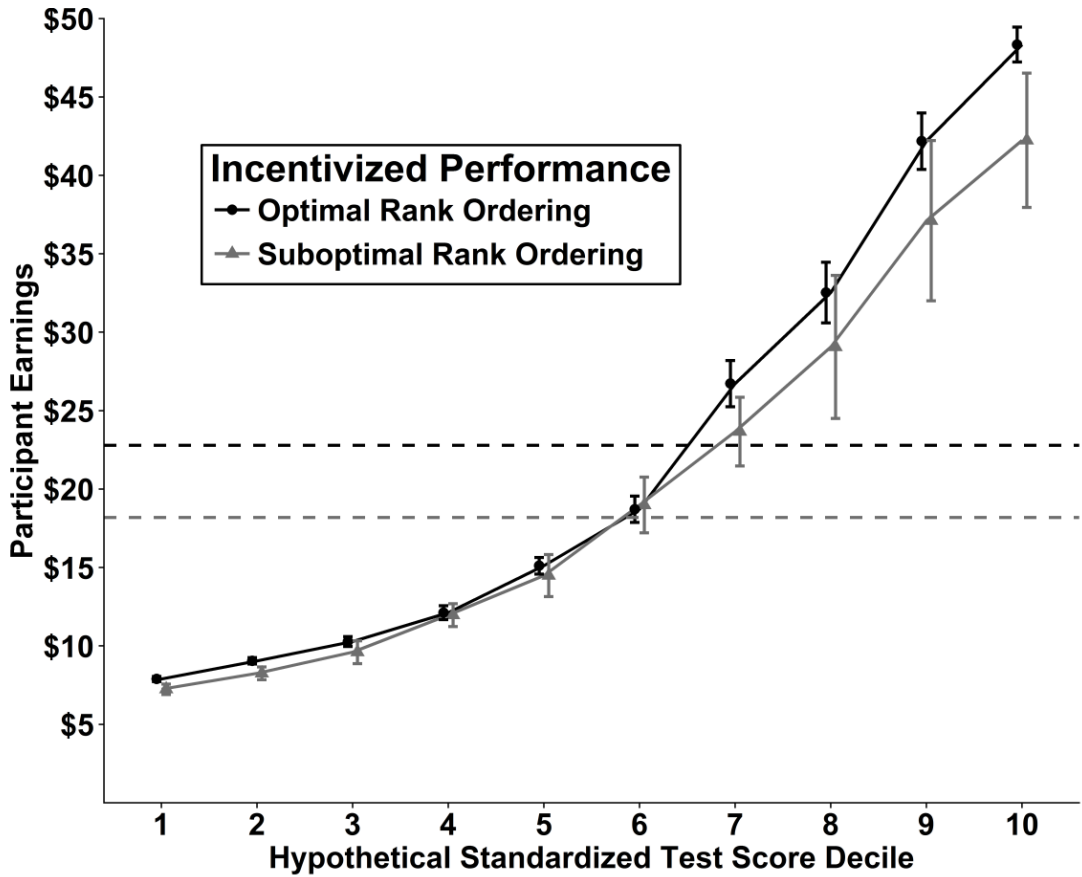

**Notes:** This figure summarizes average experimental earnings as a function of both truth-telling status (optimal versus suboptimal rank ordering) and participants' randomly assigned test scores. The dashed lines represent the overall average earnings for participants who submitted suboptimal ($18.20) and optimal ($22.80) rank orderings. The solid lines denote average earnings within each decile of assigned test scores. Vertical lines at each point show 95% confidence intervals. See SI Appendix table S5 for statistical comparisons.

**Fig. 4. Predictors of Truth-telling.**

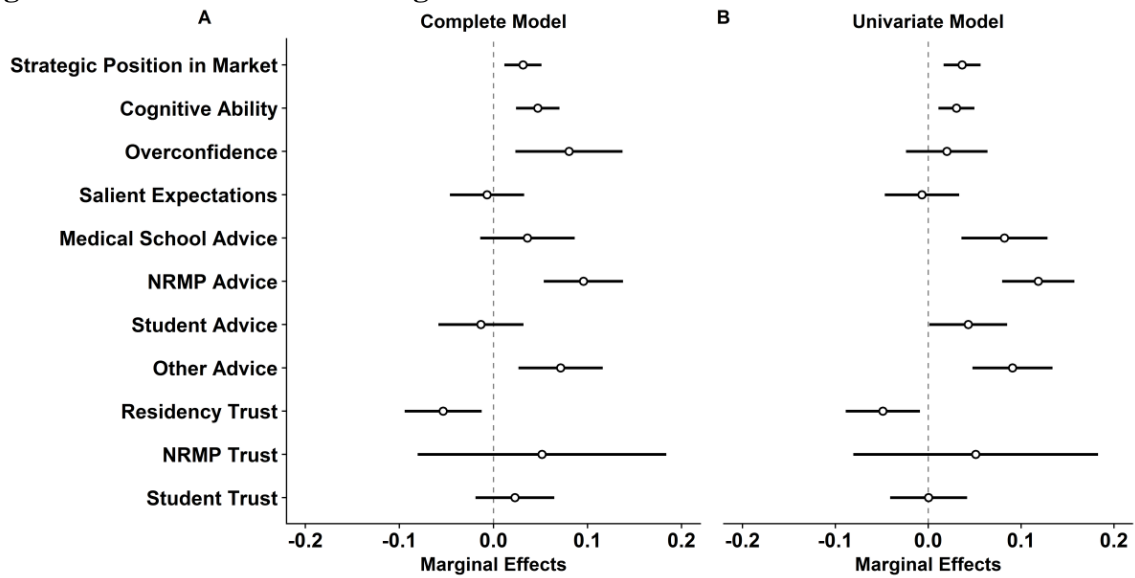

**Notes:** Plotted are estimated average marginal effects derived from a logit model predicting whether participants reported truthful preferences. To illustrate the interpretation of effect sizes, note that a marginal effect of 0.1 corresponds to a 10 percentage point increase in the rate of truthful reporting. Panel A presents estimates obtained from the complete model, including the entire battery of predictors. Panel B presents the estimate for each univariate model, predicting truth-telling with only the single variable represented in that row. Participants' HST score and Raven's performance are normalized. All other measures are binary. Horizontal lines at each data point represent 95% confidence intervals. See SI Appendix table S9 for the regression output. Sample for all regressions: 1,714.

Supporting Information for

# An Experimental Investigation of Preference Misrepresentation in the Residency Match

Alex Rees-Jones, Samuel Skowronek.

correspondence to: alre@wharton.upenn.edu

**This file includes:**

Materials and Methods
Figs. S1 to S12
Tables S1 to S9

Links to all data files will be made publicly available at the time of publication.

**Materials and Methods**

Survey Administration

In 2016, 164 medical schools were fully accredited by the Association of American Medical Colleges (AAMC). We contacted the 147 medical schools in the United States and Puerto Rico.[‡‡] On December 15th, 2016, we sent the email below to the one of the medical schools administrators (in most cases, the email was sent to the Associate Dean of Student Affairs):

> Dear [MEDICAL SCHOOL ADMINISTRATOR],
>
> I am contacting you in your capacity as the [MEDICAL SCHOOL ADMINISTRATOR'S JOB TITLE] to ask if I might recruit your graduating medical students to a study on decision making in the residency match. Our study, which I am conducting in collaboration with Professor Alex Rees-Jones at the University of Pennsylvania, is aimed to help us understand how students think about the matching process, what common mistakes are made, and how to best design the advice that schools provide to their students as they go through this process.
>
> We are working to recruit medical students nationwide and would greatly appreciate your school's participation. All we are asking from you is permission to email a link to our survey to your students in February, right after the match. The survey should only take 10-20 minutes for students to complete and provides students with a chance to win prizes as compensation for their time. All responses would be anonymous and pose minimal risk to the student participants.
>
> Please let us know if we can answer any additional questions you have over email or over the phone. If you are interested, I would be happy to share the current draft of the survey. Furthermore, all materials will be approved by our Institutional Review Board, and we would be happy to provide the relevant documentation once it is finalized. Thank you for your consideration.
>
> Sincerely,
> Melissa Beswick
> Research Coordinator
> The Wharton School

We sent the email below to medical school administrators who requested IRB documentation:

> Dear [MEDICAL SCHOOL ADMINISTRATOR],

---

[‡‡] We chose not to contact the medical schools in Canada because the majority of students graduating from Canadian medical schools do not participate in the NRMP.

> I am writing to let you know that the IRB has reviewed our protocol and has given us approval to proceed. I have attached the relevant documentation. Our survey research is anonymized and poses no tangible threat to participants, so the IRB has determined that it does not require continued IRB oversight and is thus considered exempt.
>
> Now that you have this documentation, I wanted to confirm that you are able to participate and distribute this survey to your students. If so, I will be in touch in February to send the survey link. Who should I send this link to in February? Thank you again for considering our request.
>
> Best,
> Melissa Beswick
> Research Coordinator
> The Wharton School

We sent a follow-up email to medical school administrators who had not responded to first email on January 17th, 2017. It looked very similar to the first wave of emails. This wave was sent after our study received exempt status by the University of Pennsylvania's Institutional Review Board.

> Dear [MEDICAL SCHOOL ADMINISTRATOR],
>
> I am contacting you in your capacity as the [MEDICAL SCHOOL ADMINISTRATOR'S JOB TITLE] to ask if I might recruit your graduating medical students to a study on decision making in the residency match. Our study, which I am conducting in collaboration with Professor Alex Rees-Jones at the University of Pennsylvania, is aimed to help us understand how students think about the matching process, what common mistakes are made, and how to best design the advice that schools provide to their students as they go through this process.
>
> We are working to recruit medical students nationwide and would greatly appreciate your school's participation. All we are asking from you is permission to email a link to our survey to your students in February, right after students submit their match preferences but before they receive the results. The survey should only take 10-20 minutes for students to complete and we compensate all participating students with an Amazon.com gift card valued between $5.00 and $50.00; the amount is contingent upon their survey performance. Additionally, we would be happy to send you resulting recommendations for medical advising and a copy of the academic paper.
>
> The University of Pennsylvania's Internal Review Board (IRB) has reviewed our protocol and has given us approval to proceed. Our survey research is anonymized and poses no tangible threat to participants, so the IRB has determined that it does not require continued IRB oversight and is thus considered exempt.
>
> Please let us know if we can answer any additional questions you have over email or over the phone. If you are interested, I would be happy to share the current draft of the survey or documentation from the IRB. Thank you for your consideration.

Sincerely,
Melissa Beswick
Research Coordinator
The Wharton School

Shortly after 9 PM on 2/22/17 (the deadline for medical students to submit their match preferences), we sent out the following email to the schools who had agreed to participate.

Dear [MEDICAL SCHOOL ADMINISTRATOR],

Thank you again for agreeing to let us survey your medical students participating in the match. It is important for our study to survey students as soon as possible after they submit preferences to the NRMP, so we'd greatly appreciate if you could forward our recruitment email at your earliest convenience. It is included below.

Please let me know once you have received this and contacted your students with the survey link. I would appreciate it if you could CC me on the email. I would be happy to answer any additional questions. Thank you.

Best,
Melissa Beswick
Research Coordinator
University of Pennsylvania

Dear graduating medical student,

You are being contacted because your school is participating in a study on the decision-making process of students in the NRMP match. If you participated in the 2017 NRMP match, we would greatly appreciate your participation in our anonymous, 10-minute survey. As a token of our appreciation, participants will earn an Amazon gift card voucher valued between $5.00 and $50.00; based on past versions of the study, we expect the average respondent will earn $21.00.
The survey can be accessed here: [LINK]

The results of this study will provide information on how medical students select residency programs and will assist in the advising and preparation of future generations of students. If you have any questions about this study, please contact Melissa Beswick at [ANONYMIZED].

We thank you and deeply appreciate your time and participation.

Alex Rees-Jones, University of Pennsylvania

Sam Skowronek, University of Pennsylvania
Melissa Beswick, University of Pennsylvania

On 2/28/17, we sent out a follow-up email to the schools who had not yet made the survey live.

Dear [MEDICAL SCHOOL ADMINISTRATOR],

Thank you again for agreeing to let us survey your medical students participating in the match. It is important for our study to survey students as soon as possible after they submit preferences to the NRMP, so we'd greatly appreciate if you could forward our recruitment email at your earliest convenience. It is included below. Do you think you will be able to distribute it soon?

Please let me know once you have received this and contacted your students with the survey link. I would appreciate it if you could CC me on the email. I would be happy to answer any additional questions.

Many thanks,
Melissa Beswick
Research Coordinator
The Wharton School

Dear graduating medical student,

You are being contacted because your school is participating in a study on the decision-making process of students in the NRMP match. If you participated in the 2017 NRMP match, we would greatly appreciate your participation in our anonymous, 10-minute survey. As a token of our appreciation, participants will earn an Amazon gift card voucher valued between $5.00 and $50.00; based on past versions of the study, we expect the average respondent will earn $21.00.
The survey can be accessed here: [LINK]

The results of this study will provide information on how medical students select residency programs and will assist in the advising and preparation of future generations of students. If you have any questions about this study, please contact Melissa Beswick at [ANONYMIZED].

We thank you and deeply appreciate your time and participation.

Alex Rees-Jones, University of Pennsylvania
Sam Skowronek, University of Pennsylvania
Melissa Beswick, University of Pennsylvania

Sample Exclusions

Several observations were excluded from the dataset prior to data analysis. These exclusions were made in accordance with the authors' pre-registration plan or general best practices of data cleaning.

The following observations were excluded:
1. Observations that represented participants' survey responses after already participating once.
2. Observations that represented demonstrable nefarious behavior.
3. Observations that represented participants' incomplete survey responses.
4. Observations recorded by participants who indicated that they did not participate in the 2017 NRMP match.
5. Participants who spent an insufficient amount of time on the incentivized exercise pages.
6. Observations recorded during an Amazon Web Service outage.

Table S4 presents the number of observations excluded by each participating medical school.

*Multiple observations from the same participant.* Participants answered two surveys in this experiment. The first survey contained all questions related to our hypotheses. Participants were automatically passed to a second survey with a separate url. Here, participants were informed of their match result in the incentivized simulation and provided their email address in order to receive the monetary bonus from their match. The two-survey design allowed us to assure participants that their email addresses would not be linked to their responses in the final experimental dataset. For clarity of exposition, we refer to the first survey as the experimental survey and the second survey as the reward survey.

Upon investigation, the reward survey dataset contained 40 email addresses with multiple observations; we temporarily matched these 40 email addresses to their experimental survey response and deleted duplicate observations.[§§] Thirty-four of these email addresses were observed twice and six email addresses were observed three times. In the decision rules that follow, we call participants second and third participations recorded in the reward survey "participant's duplicate observations" for parsimony.

The goal of this cleaning procedure was to exclude duplicate experimental survey observations. To do so, we first generated a variable that flagged the chronological order in which each participant who submitted one of the 40 email addresses took the survey. Using IP addresses, we excluded participants' second and third participations in the experimental survey using the following decision rules. (See fig. S2 for a graphical depiction of decision rules):

---

[§§] Duplicate email addresses recorded in the reward survey need not represent nefarious behavior nor necessitate that the same participant took the experimental survey more than once. If participants refreshed their web browser while the reward survey data was being recorded by the survey software (possibility because of a slow internet connection), their reward survey response would be recorded twice but their experimental survey response recorded only once.

1. If the IP address of participants' duplicate observations does not exist in the experimental survey, we did nothing.

2. If the IP address of participants' duplicate observations exists in the experimental survey once, we checked to see if the start date/time of the duplicate observations matched the end date/time of the corresponding observations in the experimental survey. If these timestamps could not be matched, we checked that the school that the participant matched with was constant across the duplicate responses and the corresponding observations in the experimental survey. If these criteria were met, we excluded this observation from the experimental survey. If these criteria were not met, we did nothing.

3. If the IP address of participants' duplicate responses exists in the experimental survey more than once, we matched these observation based on the criteria articulated in point two and excluded observations accordingly.

Using this procedure, we successfully identified the duplicate observations in the experimental survey and excluded them.
In total, 22 observations were excluded (Table S4, column 3).

*Nefarious behavior.* We actively monitored the completion of surveys throughout the sampling period, attempting to identify repeat-observations like those discussed above before payments were issued. On March 1, we noticed a sudden spike in survey-completion-activity from one participating medical school. Further, only five of the 43 observations collected contained a school-affiliated email address. Eight of these observations had previously been excluded based on the "Multiple observations from the same participant" criteria outlined above. Only the five observations that provided a school-affiliated email address were retained for analysis.
In total, 24 observations were excluded (Table S4, column 4), due to our strong belief that these represented attempts to "farm" the survey.

*Incomplete responses.* We excluded 640 observations because they were incomplete (Table S4, column 5).

*Non-participants of the 2017 NRMP match.* Our recruitment materials explicitly requested only participants in the 2017 NRMP match, but nevertheless we expected to have some non-participants click through into our survey. The first question participants were asked in our survey was whether they participated in the 2017 NRMP match. By design, those who indicated they did not participate were removed from the survey and did not answer any further questions.
In total, 20 observations were excluded for this reason (Table S4, column 6).

*Insufficient Time.* In accordance with our pre-registration plan, we excluded participants who spent less than 30 seconds combined on the page that explained the instructions of the hypothetical match and the page in which participants submitted their hypothetical rank ordering. This exclusion was made in order to exclude participants who do not pay attention to the information necessary to thoughtfully respond to our simulated matching exercise.

In total, 58 observations were excluded for this reason (Table S4, column 7).

*Amazon Web Service outage.* On February 28$^{th}$ at 10:45am, Amazon Web Services (AWS) crashed. This affected participants' ability to complete the survey. For example, many participants reported that they were unable to see the Raven's Matrices during this time. By 2:13pm, we closed all active surveys and did not reopen them until AWS was back online later that evening. We excluded all observations recorded between 10:45am and 3:00 pm on February 28$^{th}$ due to the high occurrence of survey errors during this time window.

In total 159 observations were excluded (Table S4, column 8).

Free Response Analysis

When participants indicated which sources provided them with advice about their NRMP submission, we asked them to explain what advice they received from each. We placed no restrictions or requirements on participants' responses; therefore, not all participants who received advice from a given source wrote a response. In total, we collected 2,960 free-responses, where 76.2% of our sample wrote at least one free response. The average number of words per response was 14.22 (SD=14.28).

We developed 36 items to code each free response (See table S7). Items 1-13 identified what advice participants received, items 14-22 identified who (or what) provided advice to participants and items 23-36 identified through what medium(s) advice was conveyed.

Four research assistants (RAs) independently coded responses on all 36 items, and each response was coded by two RAs. All items were coded on a binary scale where the free response was given the value one if the participant explicitly mentioned the content of the item and zero if not. When the RAs disagreed on an item, a fifth RA was brought in to break the tie. All RAs were blind to hypotheses.

## Survey Screenshots

**Residency Matching Survey**

We have contacted you because administrators in your medical program have agreed to let us survey your program's current seniors in an effort to better understand how residency applicants form their NRMP rankings. Your participation is voluntary and is greatly appreciated. Your participation will help inform our research on the experience of medical students as they go through this important process and hence may benefit other medical students and medical programs in the future.

Throughout this survey, you will be asked questions about your NRMP preference rankings and how you evaluate your chosen residencies. It is important for our study that you complete this survey on your own and answer truthfully. You may withdraw from the study at any time.

If you agree to participate, please fill out this survey without discussing it with others. We anticipate that this survey will take 10 minutes to complete.

**Contact information**: This study is being conducted by Alex Rees-Jones and Sam Skowronek at the University of Pennsylvania. If you have any questions or comments, please contact Sam Skowronek (samsko@wharton.upenn.edu). In order to guarantee the responsible conduct of research with human participants, this study has been reviewed by the University of Pennsylvania Institutional Review Board (IRB). Approval for this study was obtained on 1/3/2017, filed under protocol number 826627. Comments, concerns, or complaints about this study can be filed with the IRB by calling (215) 573-2540.

**Payment for Participation**: Participants will earn an Amazon gift card voucher valued between $5.00 and $50.00. Based on past versions of this study, we expect the average respondent to earn $21.00.

**Eligibility:** To be eligible to participate in this survey, you must be a medical student participating in the 2017 NRMP match.

If you agree to fill out this survey, please click the ">>" button below to begin.

Are you participating in the 2017 NRMP match?

○ Yes
○ No

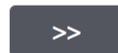

In this exercise, you will go through a simulated matching process much like the NRMP match. You will attempt to match to one of five hypothetical residency programs, and the payment you receive for taking this survey will depend on where you match. We will apply the standard algorithm that was used by the NRMP; as a reminder, an example of how this algorithm works is available here.

There are five residency programs, each with 10 positions available. You are one of 50 students applying to them. All students agree on the ranking of these programs, from best to worst.

These residency programs make their decisions based on several factors. One important factor is students' performance on the "Hypothetical Standardized Test" (HST). For this exercise, imagine that you took the HST and you scored in the 58 percentile. This means that 58% of the medical students looking to enroll in the same programs scored worse than you and 42% scored as well as or better than you.

Below is the rank ordering of residency programs, the average HST score of their admitted students, and the value of the Amazon.com gift card that you will receive if you match to that program.

| Rank | Residency program | Average HST percentile score of admitted students | Amazon.com gift card value |
|---|---|---|---|
| 1. | Maplecrest | $80^{th}$ percentile | $50.00 |
| 2. | Birch Hill | $65^{th}$ percentile | $25.00 |
| 3. | Elm South | $50^{th}$ percentile | $15.00 |
| 4. | Hickory Bridge | $35^{th}$ percentile | $10.00 |
| 5. | Pine Peak | $20^{th}$ percentile | $7.50 |

It is also possible that you will not match to a program in which case you will receive a $5.00 Amazon.com gift card.

Which residency program do you expect to match with?

Residency Program

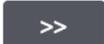

If the participant clicked on the link to the NRMP website, they were directed to the example presented on the NRMP website that illustrates, using 4 medical students and 3 residencies, how matches are determined. All participants proceeded to the following page.

Participant who were <u>not randomized to</u> indicate the residency where they expected to match saw the following page:

Based on what you know about the NRMP's matching algorithm, please list the order in which you would like to apply to these residency programs. We will run the matching algorithm, assign you to a program, and provide you an Amazon.com gift card based on which program you match with at the end of this survey.

As a reminder, below is your HST score, the list of residency programs, and the bonus associated with each one.

<u>Your HST score: 72 percentile</u>

| Rank | Residency program | Average HST percentile score of admitted students | Amazon.com gift card value |
|---|---|---|---|
| 1. | Maplecrest | 80th percentile | $50.00 |
| 2. | Birch Hill | 65th percentile | $25.00 |
| 3. | Elm South | 50th percentile | $15.00 |
| 4. | Hickory Bridge | 35th percentile | $10.00 |
| 5. | Pine Peak | 20th percentile | $7.50 |

You must apply to at least one program, but you are not required to apply to every program; you may leave blank application opportunities that you choose to forego.

Please submit your applications below:

First Application  
Second Application  
Third Application  
Fourth Application  
Fifth Application

Participants who were randomized to indicate the residency where they expected to match saw the following page:

Based on what you know about the NRMP's matching algorithm, please list the order in which you would like to apply to these residency programs. We will run the matching algorithm, assign you to a program, and provide you an Amazon.com gift card voucher based on which program you match with at the end of this survey.

As a reminder, below is your HST score, the list of residency programs, the bonus associated with each one, and the university you expect to match with.

Your HST score: 38 percentile

You indicated that you expect to match with Birch Hill

| Rank | Residency program | Average SCT percentile score of admitted students | Amazon.com gift card value |
|---|---|---|---|
| 1. | Maplecrest | 80th percentile | $50.00 |
| 2. | Birch Hill | 65th percentile | $25.00 |
| 3. | Elm South | 50th percentile | $15.00 |
| 4. | Hickory Bridge | 35th percentile | $10.00 |
| 5. | Pine Peak | 20th percentile | $7.50 |

You must apply to at least one program, but you are not required to apply to every program; you may leave blank application opportunities that you choose to forego.

Please submit your applications below:

First Application
Second Application
Third Application
Fourth Application
Fifth Application

All participants then proceeded to the following pages.

Now, we will ask you a series of questions about your real submission to the NRMP.

How many residency programs did you list in your submission to the NRMP?

[ ▼ ]

---

When forming the ranking of residencies to submit to the NRMP, some candidates submit an ordering that is not the true order of how desirable they find the programs.

When forming your list, did you report the exact ordering of your true preferences?

○ Yes
○ No

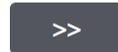

You indicated that you did not report the exact ordering of your true preferences.

Please indicate your true preference ordering over the programs that you ranked. Click and drag the boxes below until they appear in order of your true preferences. List your true favorite program first, your true second favorite program second, etc.

Your #1 NRMP Choice
Your #2 NRMP Choice
Your #3 NRMP Choice

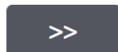

When you submitted your rank ordering to the NRMP, which program did you expect to match with?

○ Your #1 NRMP Choice
○ Your #2 NRMP Choice
○ Your #3 NRMP Choice
○ You did not expect to match with any program.

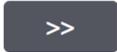

How confident are you that you will be matched with Your #2 NRMP Choice?

| 0% confident | 25% confident | 50% confident | 75% confident | 100% confident |

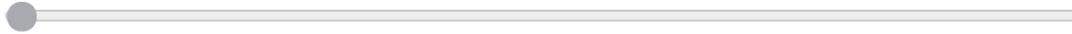

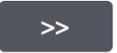

Did you receive advice about how to rank order residency programs before submitting them to the NRMP?

- ● Yes
- ○ No

Where did you receive this advice?
(Check all that apply)

- ☐ From current and/or past medical students who participated in the NRMP.
- ☐ From your medical school.
- ☐ From the NRMP website.
- ☐ From other sources not listed above.

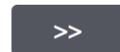

Do you trust the advice you were provided by current and/or past students?

○ Yes
○ No

What advice were you given by current and/or past students?

[ ]

You indicated that you received advice regarding how to submit your preferences to the NRMP from sources other than the NRMP website, current and past medical students, and your medical school's advisor.

Do you trust this advice?

○ Yes
○ No

Who gave you this advice, and what advice did they give you?

[ ]

Please click and drag the boxes below so that the order reflects how influential each piece of advice was for submitting your NRMP rank ordered list.

The advice that was **most** influential should be at the **top** of the list.
The advice that was **least** influential should be at the **bottom** of the list.

    The advice from current and/or past medical students

    The advice you received from sources other than the NRMP website, you medical school, and current and/or past medical students

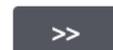

Do you trust that all of your ranked residency programs ranked medical students based on truthful assessments of their quality?

○ Yes
○ No

Do you trust that all other medical students submitted a truthful rank ordering of their preferences to the NRMP?

○ Yes
○ No

Do you trust the NRMP to run the matching algorithm honestly?

○ Yes
○ No

>>

**On the next page you will see a series of problems that are taken from a test of spatial reasoning.**

**Please solve as many of these problems as you can <u>in 5 minutes</u> by clicking the number of the choice that completes the pattern.**

**There is a timer at the top of the page for your reference.**

>>

[We used seven matrices (matrices 1, 2, 3, 4, 26, 29, and 31) from the Advanced Progressive Matrices Set II booklet for this study. Due to copyright laws, we are unable to post these specific matrices. Below is a sample of a Raven's matrix. This matrix was not used in our study. To see the matrices used for this study, you will need to purchase the Raven's Advanced Progressive Matrices Kit published by Pearson Education.]

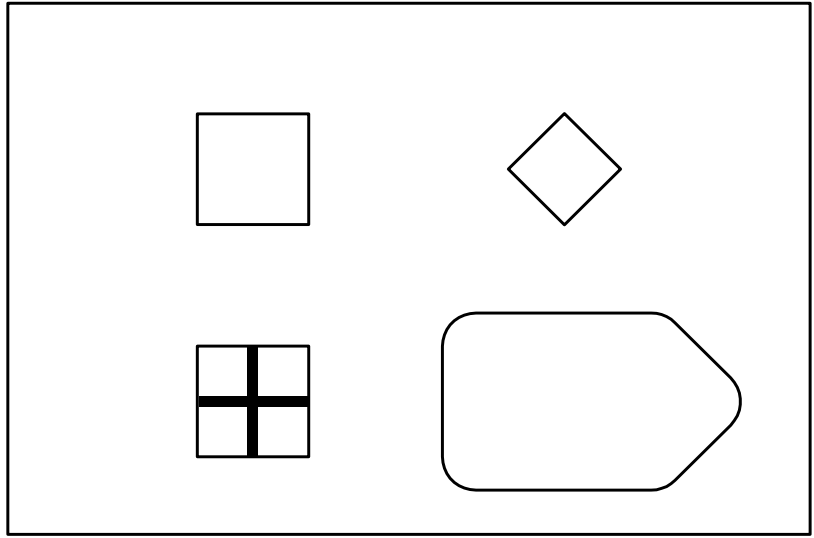

1
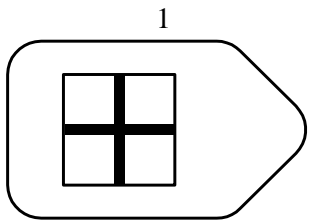

2
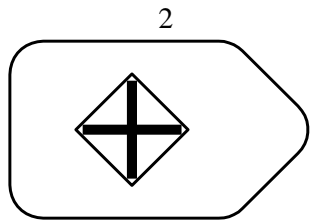

3
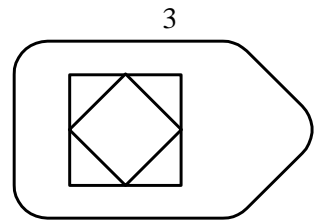

4
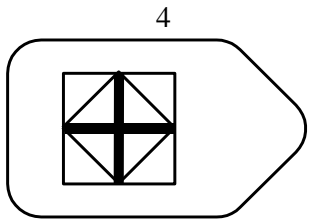

5
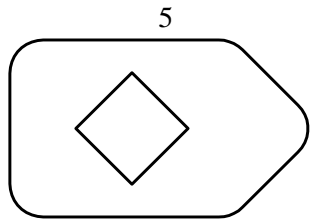

6
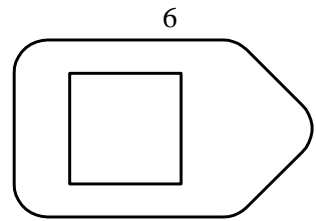

Please let us know how you think you performed on the spatial reasoning task you just finished compared to other medical students taking this survey by completing the following statement using the slider scale.

I think I did better than _____ % of other medical students taking this survey on the spatial reasoning tasks above.

| 0 | 10 | 20 | 30 | 40 | 50 | 60 | 70 | 80 | 90 | 100 |

%

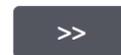

Now, just a few questions about yourself:

What year did you take the MCAT?

[ ▼ ]

What was your MCAT score?

[ ▼ ]

In the year that you took the MCAT, what percentage of MCAT takers do you estimate scored lower than you?

*Please make your best guess without consulting other sources. This does not need to be completely accurate.*

% of MCAT takers who received a lower score than you

0    10    20    30    40    50    60    70    80    90    100

[                                                              ]

What is your specialty or specialties? Select the specialties of any programs you applied for in the NRMP match from the dropdown menu.

| | |
|---|---|
| First Specialty | [ ▼ ] |
| Second Specialty | [ ▼ ] |
| Third Specialty | [ ▼ ] |
| Fourth Specialty | [ ▼ ] |

Did you participate in the NRMP "Couples Match?"

○ Yes
○ No

How well do you think you understood the NRMP matching algorithm when you submitted your rank ordering to the NRMP?

| Not well at all | Slightly well | Moderately well | Very well | Extremely Well |
|---|---|---|---|---|
| ○ | ○ | ○ | ○ | ○ |

How well do you think you understood the NRMP matching algorithm <u>compared to other medical students</u> taking this survey?

| I understood the algorithm <u>much worse</u> than others | I understood the algorithm <u>slightly worse</u> than others | I understood the algorithm <u>about the same</u> as others | I understood the algorithm <u>slightly better</u> than others | I understood the algorithm <u>much better</u> than others |
|---|---|---|---|---|
| ○ | ○ | ○ | ○ | ○ |

What is your current age?

[           ]

What is your gender?

○ Male
○ Female
○ Other [           ]

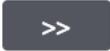

We have run the matching algorithm, and you have matched with the Birch Hill Residency Program. In order to receive your $25.00 Amazon.com gift card, please enter your **school affiliated email address** below. Your Amazon.com gift card will be sent to this email address within five business days.

○ Email Address:

[ ]

○ I choose to not provide my email address and forego receiving the gift card.

---

Preserving the anonymity of your responses is important to us. You have been passed to a different web survey to collect your email information for payment. Through this design, the email address you enter above is recorded in a separate database from that recording survey responses. It will not be linked to your individual survey responses, and it will solely be used for financial reporting requirements associated with providing these payments.

Please click the >> button to complete this survey.

>>

---

**Thank you for completing this survey. We greatly appreciate your participation. Your Amazon.com gift card will be sent to the email address that you provided within five business days.**

**You may contact Sam Skowronek at samsko@wharton.upenn.edu with comments, questions, or concerns about this survey.**

**You may close this webpage.**

Simulation of Match Results:

Final matches in our experiment were determined by combining the participants' reported preferences with those of 49 simulated students and 5 simulated residency programs.

Preferences held by other students were simulated assuming optimal preference reporting.

Preferences held by residency programs were simulated as follows. We assigned a random valuation for each of the 49 simulated students, drawn independently and uniformly from the intergers spanning 0 to 99, inclusive. The valuation for the participant was set to be their HST score. Preferences were then formed by sorting this valuation number from largest to smallest. The same preference order applied to all five residency programs.

This simulation of residency preferences may be thought of in the context of a random utility model. Consider the case where valuations are determined by combining HST scores with a random error term, meant to capture other valued features of an application. Denote this valuation of student i as $V_i^1 = HST_i + \varepsilon_i$. Since preference orderings are invariant to adding a constant to the utility function, we may equivalently rationalize residency preferences with the valuation function $V_i^2 = HST_i + \varepsilon_i - \varepsilon_s$, where $\varepsilon_s$ is the random error draw for the non-simulated match participant. Our preference simulation procedure can be interpreted as sorting on values of $V_i^2$, where the random draws assigned to the 49 simulated students correspond to draws from the distribution of $HST_i + \varepsilon_i - \varepsilon_s$. We randomly sample this final value, rather than the individual components, to minimize the computation necessary in each run of the simulation.

With these estimates in hand, and assumed class-sizes of 10 for each program, we run the deferred acceptance algorithm as described in (15).

Preregistration Document

**As Predicted: "Rees-Jones Skowronek NRMP Study" (#3109)**
**Created:** 02/21/2017 02:21 PM (PT)

**Author(s)**
Alexander Rees-Jones (Wharton) - alre@wharton.upenn.edu
Samuel Skowronek (Wharton) - samsko@wharton.upenn.edu

**1) What's the main question being asked or hypothesis being tested in this study?**
A great deal of care has gone into designing matching mechanisms that make it incentive compatible for participants to truthfully report their preferences. The National Residency Match Program (NRMP), which matches nearly all graduating medical students in North America to residency programs, is arguably the most important application of this mechanism. However, recent work (e.g., Rees-Jones 2016) suggests that at least some students try to "game the system" in these settings, despite the futility of such efforts. This project aims to better understand why medical students misrepresent their preferences in matching mechanisms. We test whether:

H1) Expectations: That propensity to misrepresent preferences is driven by a desire to match to the residency program that the student expected to match with ex ante (leading students to list an attainable option as their "first choice").

H2) Advice: That propensity to misrepresent preferences is driven by the advice (or lack thereof) the student received from his/her medical school, other medical students, and the NRMP website.

H3) Competition in match: That propensity to misrepresent preferences is driven by a perceived need to seek a competitive advantage when a student knows they are an undesirable match participant (e.g., due to low standardize test scores).

H4) Reasoning ability: That propensity to misrepresent preferences is driven by differences in logical reasoning ability.

H5) Overconfidence: That propensity to misrepresent preferences is driven by overconfidence.

H6) Trust: That propensity to misrepresent preferences is driven by mistrust of other market participants.

**2) Describe the key dependent variable(s) specifying how they will be measured.**
In our study, we have one primary and one secondary dependent variable.

In this study, participants are presented with a hypothetical residency matching exercise. They could match with one of five programs, and their compensation for the survey depends on which program they match to. They are then asked to submit their preference ordering over the programs, which we input to the matching mechanism. Our primary DV is an indicator of whether or not the participant submitted a "truthful" ranking of the schools. If they ranked the programs in order of their compensation, we code it as truthful. If they did not, we code it as nontruthful.

In a later section of our study, participants are asked whether or not the rank-ordered lists (ROL) that they submitted to the NRMP is an exact ordering of their true preferences. In contrast to our primary measure above, which is an incentivized measure of understanding, this measure is merely an unincentivized self-report. We therefore view our experimental measure as our DV of primary interest; however, we collect this secondary DV to assist in validating those results.

**3) How many and which conditions will participants be assigned to?**
Participants will be randomly assigned to either the expectation or the no-expectation condition.

Participants assigned to the expectation condition will be asked, in the hypothetical matching exercise, to state what residency program they expect to match with before they rank order their preferences. Participants assigned to the no-expectation condition with not make this prediction before they rank order their preferences. This provides us with a means to test H1.

Furthermore, all participants will be randomly assigned a standardized test percentile score (1-99). They are told that this is the test score they would submit to the programs in the hypothetical matching exercise, providing us with a means to test H3.

**4) Specify exactly which analyses you will conduct to examine the main question/hypothesis.**
We will test each of the 6 hypotheses described above by examining differences in the rate of truthful reporting in the primary and secondary dependent variables. The tests of each hypothesis are:

H1: Expectations
• Assessing difference in DV across the expectation condition. (Preferred test: differences in proportions)

H2: Advice
• Assessing difference in DV depending on the presence of each measured source of advice. (Preferred test: logit predicting truthful reporting on self-reported receipt of advice from other students, participants' medical school, the NRMP, and other sources) .

H3: Competition in match
• Assessing difference in DV depending on random assignment to the hypothetical test score. (Preferred test: logit predicting truthful reporting with continuous test score)

H4: Reasoning ability

• Assessing difference in DV depending on differences in logical reasoning measured by Raven's Matrices. (Preferred test: logit predicting truthful reporting with number of correct Raven's Matrices).

H5: Overconfidence
• Assessing difference in DV depending on differences in overconfidence measured by the participants' forecast of their performance on the Raven's Matrices compared to others. (Preferred test: two sample difference of proportions, split by whether or not the participant over- or under- estimated his performance)

H6: Trust
• Assessing difference in DV depending on provision of advice from all measured sources. (Preferred test: logit predicting truthful reporting based on self-reported trust of the NRMP, other students, and residency programs).

In addition to these tests which individually assess each of the hypotheses above, we will run analyses where we predict our primary DV with all measures described above. Our preferred specification will be logit regression, although we will run OLS regressions as robustness checks.

We will also test the correlation between the primary and secondary DVs.

**5) Any secondary analyses?**
• We will compare the propensity to expect to match to the first-ranked school submitted to the NRMP, split by self-reported truth-telling status. (Auxillary test of H1: Expectations)
• Assessing difference in DV depending on MCAT scores. Note that this conflates two things that are disentangled in our experimental design: underlying "ability" as well as one's competitive position in the market. (Preferred test: logit predicting truthful reporting with continuous test score) (Auxillary test of H3: Competition in match and H4: Reasoning ability)
• Assessing difference in DV depending on differences in overconfidence measured by participants' estimate of their MCAT performance relative to others who took the MCAT that year. (Preferred test: two sample difference of proportions, split by whether or not the participant over- or under- estimated his performance) (Auxillary test of H5: Overconfidence)
• We will test for gender differences in propensity to misrepresent preferences. The small amount of existing research on the topic suggests that men are more likely to misrepresent their preferences in this environment than women.

**6) How many observations will be collected or what will determine sample size? No need to justify decision, but be precise about <u>exactly</u> how the number will be determined.**
We are recruiting from a sample of approximately 4,000 medical students. Past research in this area has recruited approximately 20% of similar samples. We therefore expect approximately 800 participants. However, we will make use of any responses received during our sampling window.

**7) Anything else you would like to pre-register?**
**(e.g., data exclusions, variables collected for exploratory purposes, unusual analyses planned?)**
We are collecting data on participants' age, but do not have any ex ante hypotheses about this component. It is collected for standard demographic reporting.

We will exclude participants who spend less than 30 seconds combined on the page that explains the instructions of the hypothetical match and the page in which participants submit their hypothetical rank ordering. This exclusion is made in order to exclude participants who do not pay attention to the information necessary to thoughtfully answer our primary dependent variable.

**8) Have any data been collected for this study already?**
No, no data have been collected for this study yet

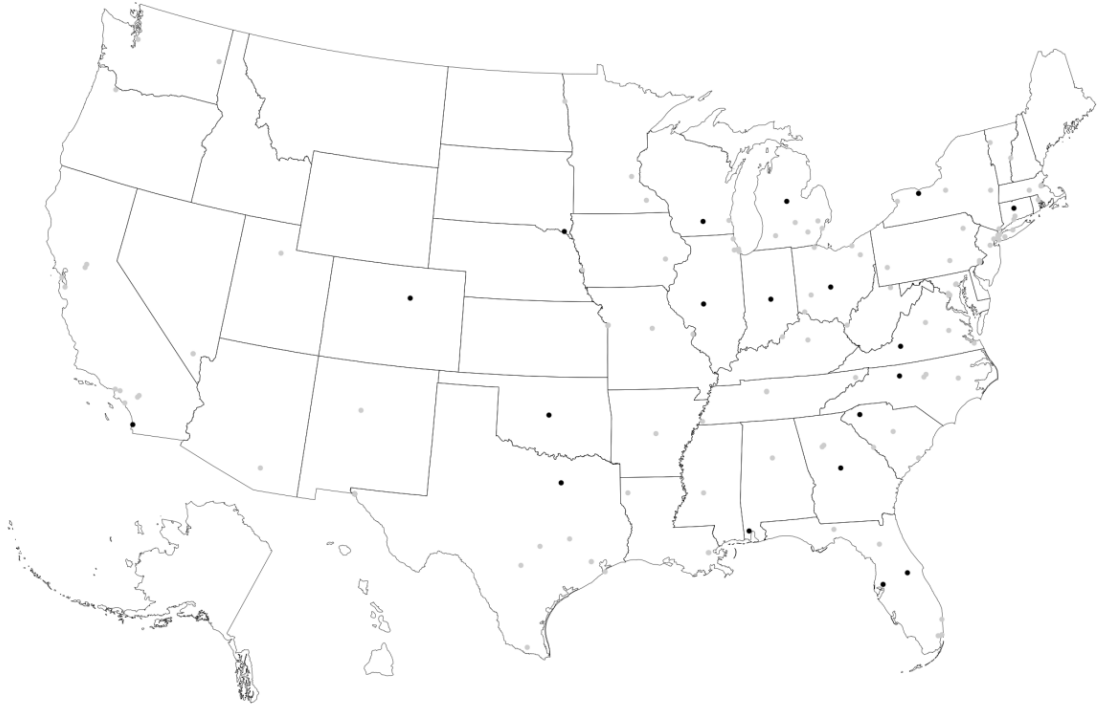

**Fig. S1. Geographic Location of Participating and Non-participating Medical Schools.** Circles represent the 143 U.S. medical schools with full accreditation from the Association of American Medical Colleges at the time of our study. Circles in black represent medical schools that participated in our study. Circles in grey represent medical schools that did not participate in our study.

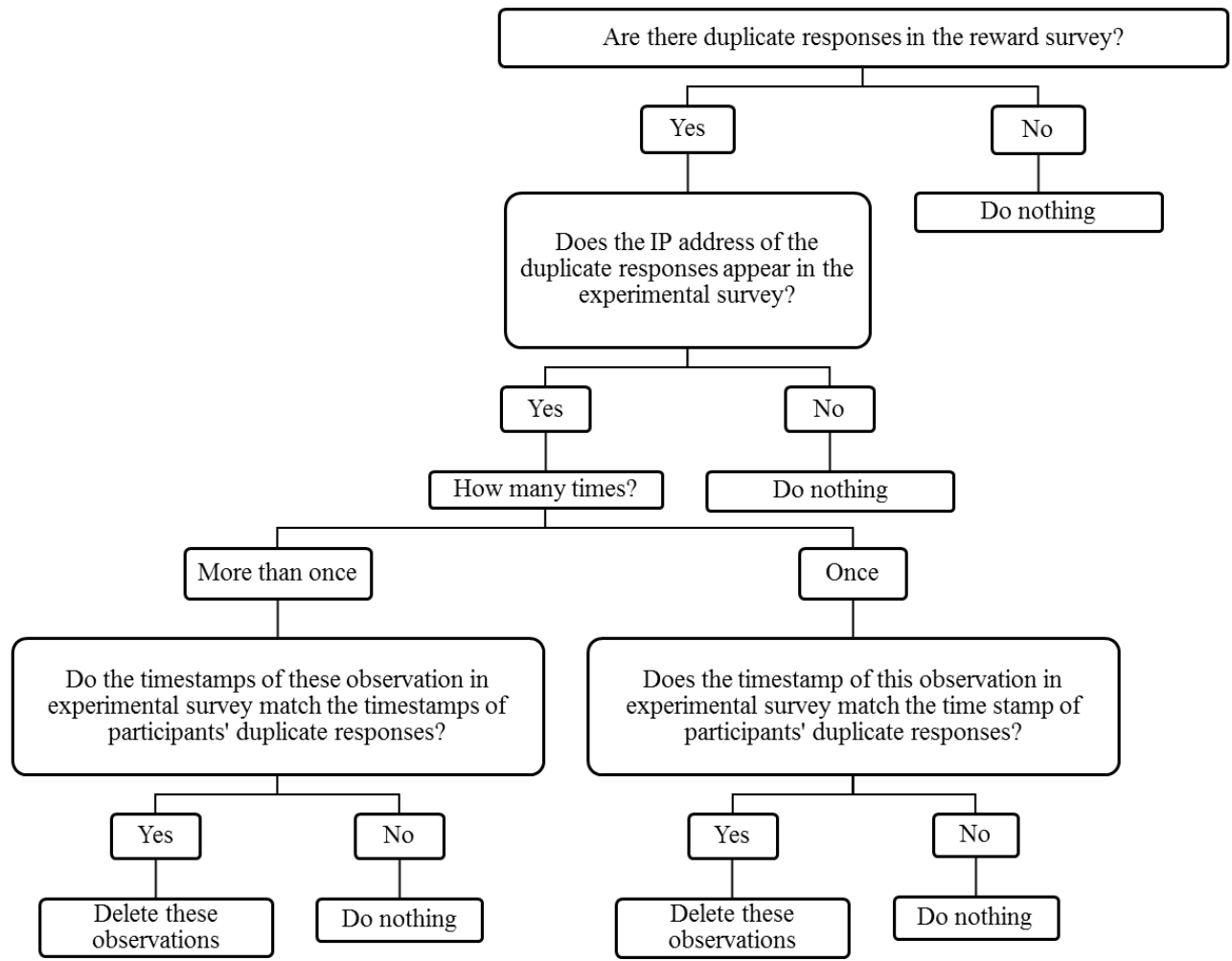

**Fig. S2. Decision Rules for Excluding Multiple Observations from the Same Participant.**

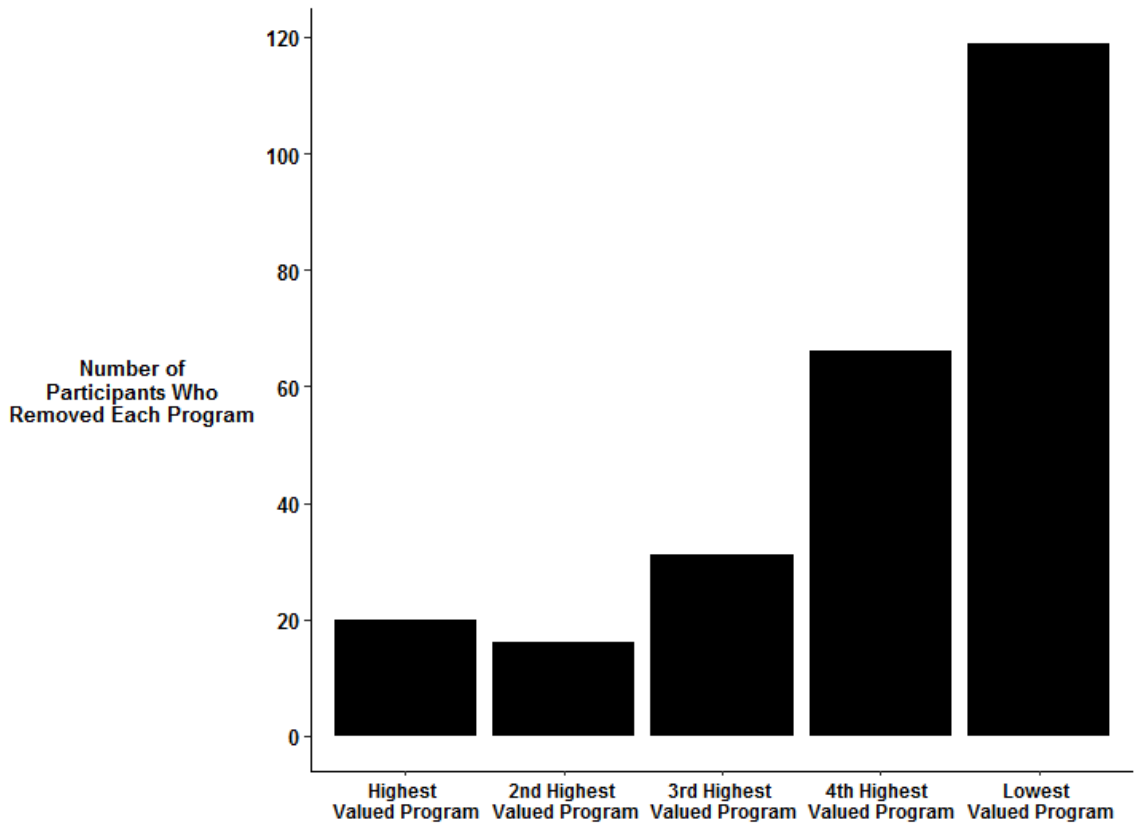

**Fig. S3. Frequency of Missing Programs in Shortened Submissions.** Shown is a histogram of the programs participants removed conditional on submitting a shortened rank-order list (N=141) in the incentivized measure of performance. Bars show the number of participants that removed each program from their ROL. For example, 20 participants who shortened their preferences removed the highest valued program from their rank-order list. These participants may also have removed other programs as well—a participant who dropped multiple programs will appear in multiple bins.

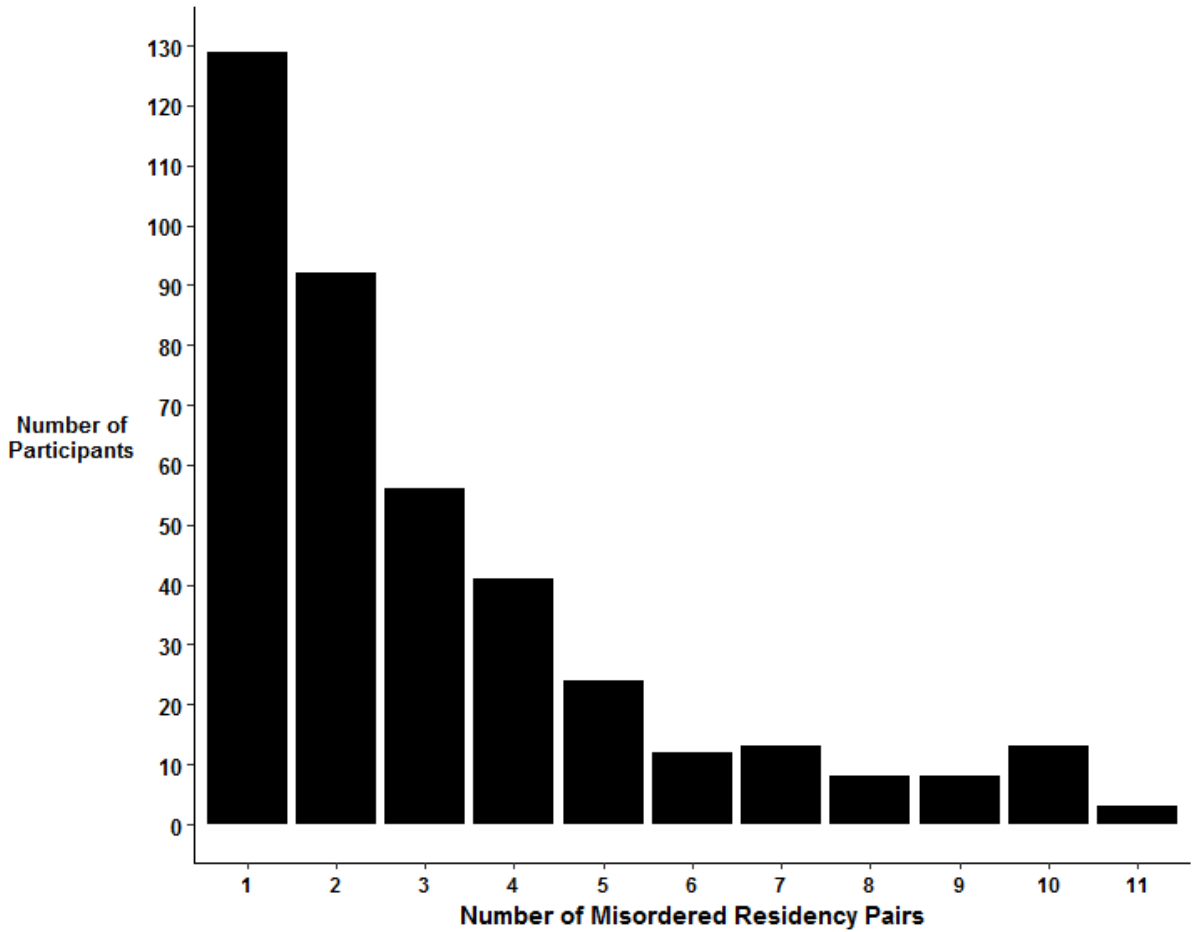

**Fig. S4. The Number of Misordered Residency Pairs Submitted During Suboptimal Play.** Shown is a histogram of the number of misordered pairs in each participants' ROL. All participants who misrepresented their preferences (N=399) are presented. For each participant, we consider each binary comparison of programs inherent in their preference list (e.g., comparing program 1 to program 2, comparing program 1 to program 3, etc.), and count the number of pairwise rankings misordered in terms of compensation.

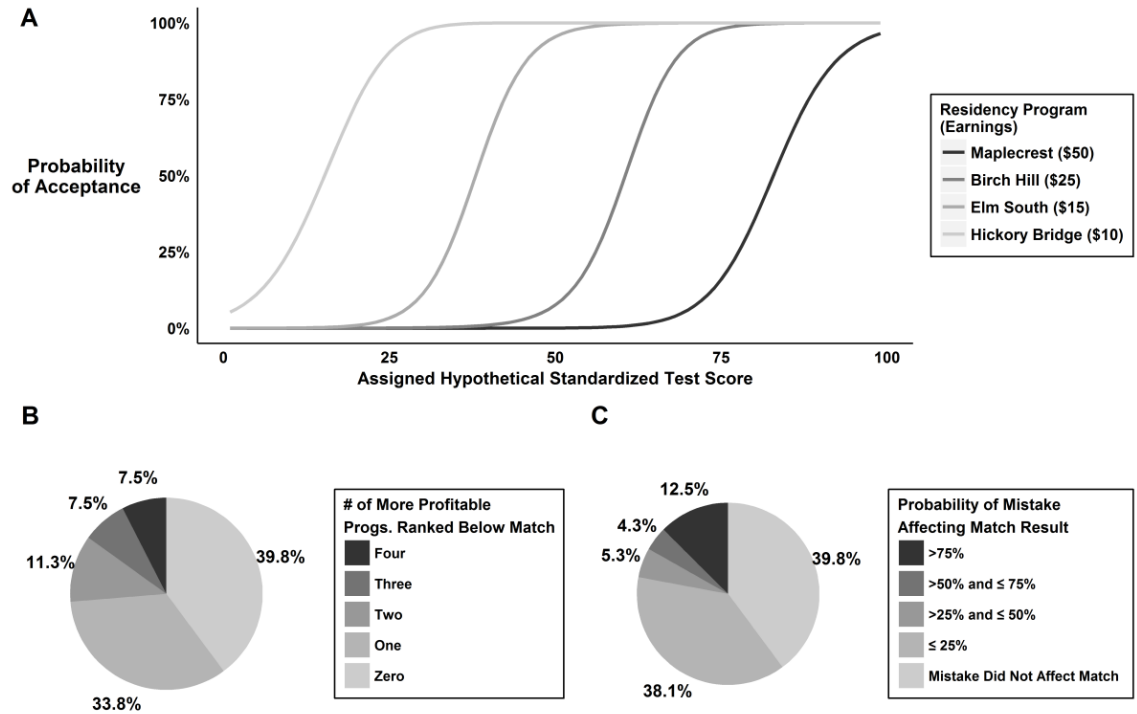

**Fig. S5. An Individual-Level Analysis of the Consequence of Suboptimal Play.**
Shown are three figures that illustrate how suboptimal play changed the match results at the individual level. The pie charts in Panels B and C present data for participants who submitted suboptimal ROLs (N=399). Panel B shows the number of more profitable programs ranked below participants' ultimate match—each of which creates a foregone opportunity for higher earnings in the experiment. To illustrate by example, participants who matched to Elm South ($15) could only lose money by misrepresentation if they ranked the $25 or the $50 program below Elm South. In Panel B we tally the number of these foregone options for each respondent who misrepresented preferences. Of course, given a participants' HST score, different foregone options have different probabilities of being obtainable: a low-HST applicant is unlikely to be admitted by the best program, and thus failing to rank it has a relatively low chance of being consequential. Panel C represents the probability that each respondent faced of matching to one of their foregone options. Estimates are derived from four logit models (presented in panel A), each predicting if a truthtelling respondent could have been admitted to each program based on a cubic spline over HST scores. Note that all respondents have a 100% probability of being admissible by the omitted, lowest-paying program.

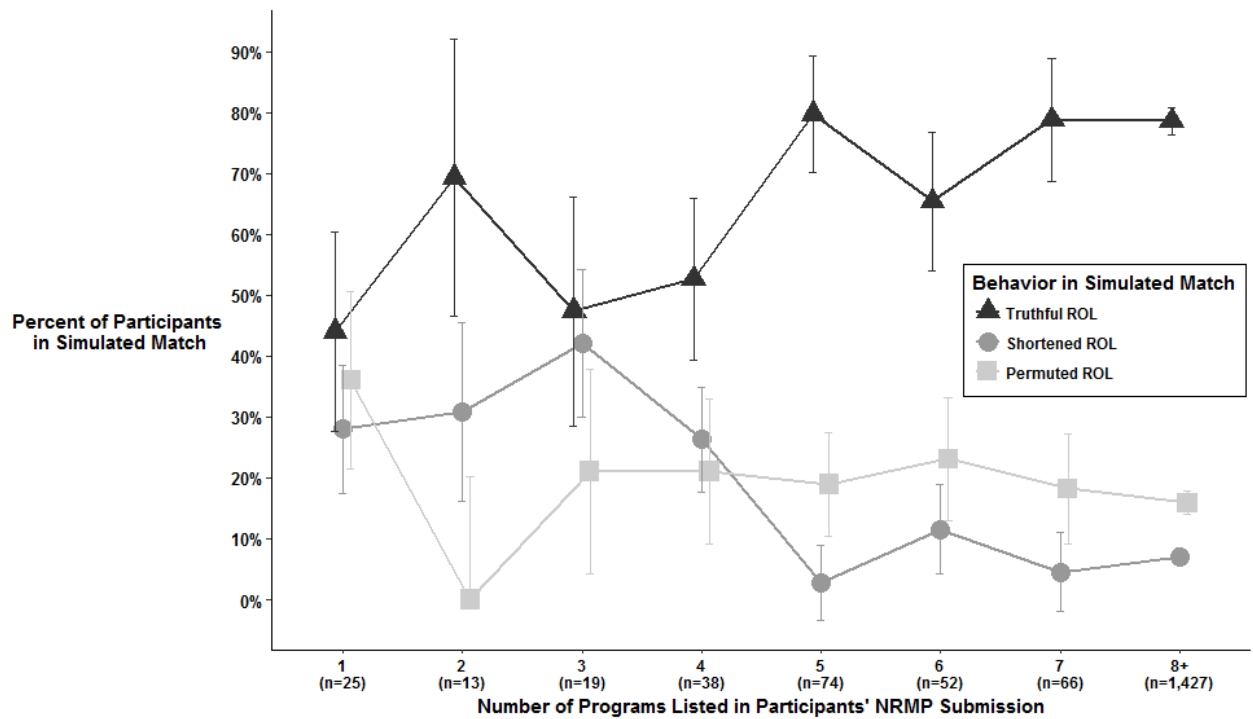

**Fig. S6. Experimental Behavior by Length of NRMP Submission.** This figure plots the propensity to submit truthful, permuted, or shortened ROLs in our simulated residency match, as a function of the number of programs listed in the participant's real NRMP submission. Plotted are the results of three separate OLS regressions in which the number of programs that participants ranked in their NRMP submission is the sole predictor. We regressed this categorical variable on participants' likelihood to tell the truth, to shorten their ROL, and to permute their ROL in the simulated match. The triangle-pointed line shows the percent of participants who told the truth in the simulated match at each length of participants' NRMP ROL. The circle-pointed line and the square-pointed line show the percent of participants who shortened and permuted their preferences, respectively. In support of the external validity of our experimental measure, participants who shortened their ROL in the NRMP submission are more likely to shorten their ROL in the simulated match. Participants who both permuted and shortened their preferences (1.6% of total sample) are represented in both the circle-pointed and square-pointed lines. Vertical lines at each data point show 95% confidence intervals.

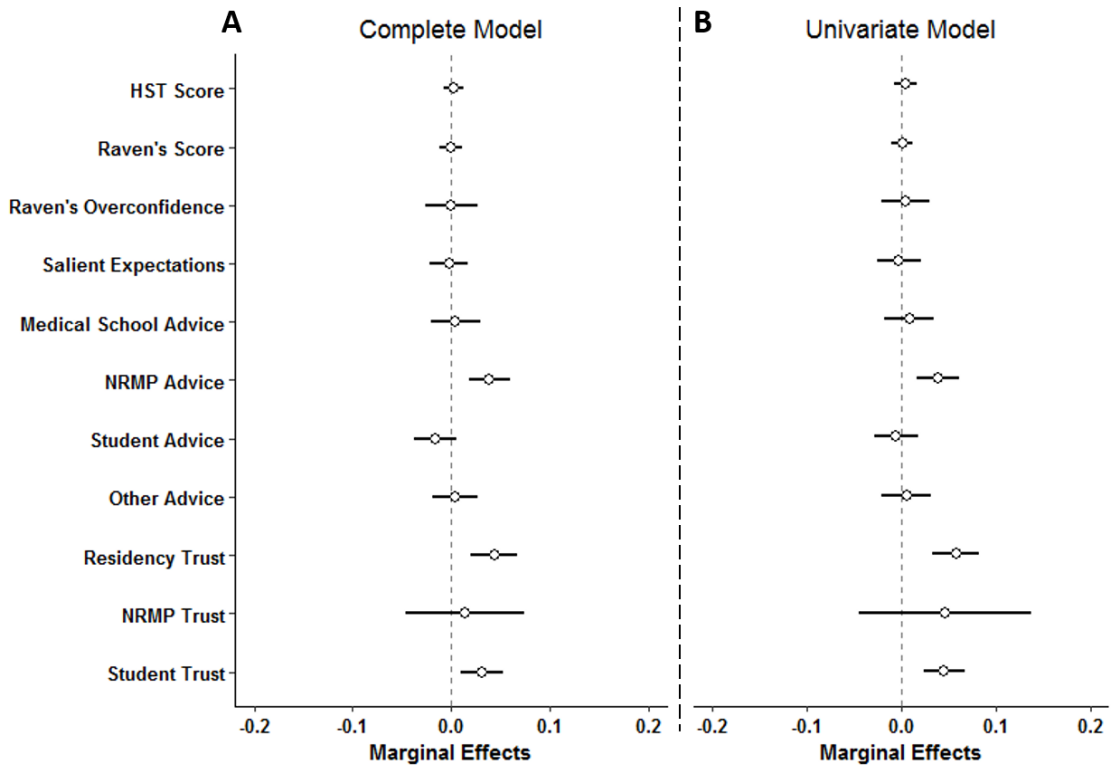

**Fig. S7. Predictors of Self-reported Truth-telling Status.** Plotted are estimated average marginal effects derived from a logit model predicting participants' self-reported truthful preferences to the NRMP. Panel A presents estimates obtained from the complete model, including the entire battery of predictors. Panel B presents the estimate for each univariate model, predicting self-reported truth-telling with only the single variable represented in that row. Participants' HST score and Raven's performance are normalized. All other measures are binary. For full descriptions of each predictor, see section II.B of the paper. Horizontal lines at each data point show 95% confidence.

Self-reported truthtelling status is measured by the yes/no response to the survey question: "When forming the ranking of residencies to submit to the NRMP, some candidates submit an ordering that is not the true order of how desirable they find the programs. When forming your list, did you report the exact ordering of your true preferences?" For comparability across studies, this phrasing was drawn from earlier work documenting self-reported misrepresentation in the match (11). This previous work argues that self-reports are likely to underestimate the rate of misrepresentation due to social-desirability biases, and documents that some respondents interpret the term "true preferences" in a manner inconsistent with economists usage of the term. Interpretation of the above results should be conducted with these caveats in mind.

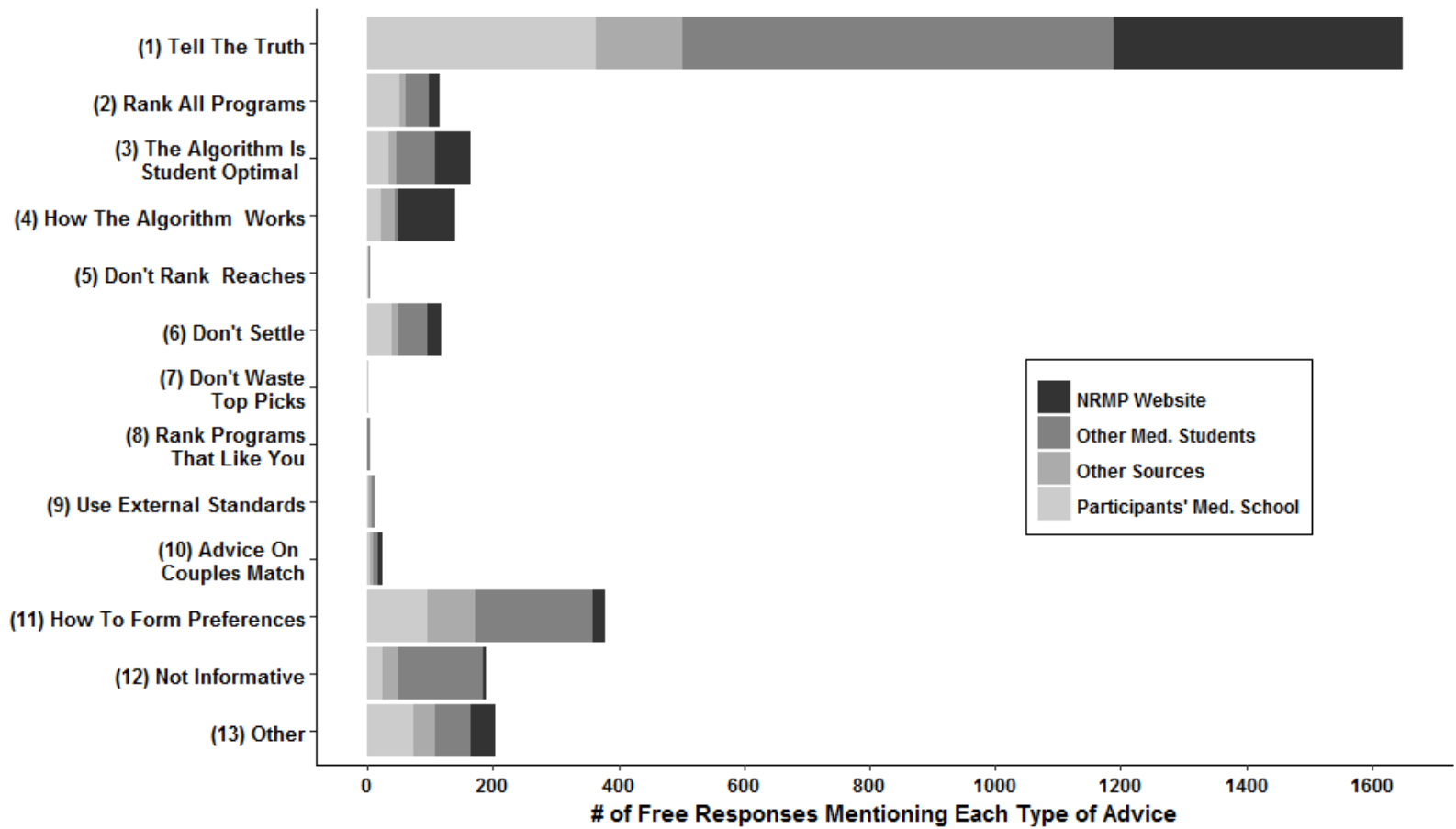

**Fig. S8. The Types of Advice Participants Received.** Participants' free responses were coded to detect 13 different types of advice. Presented are the frequencies of each type of advice participants received by the source that provided the advice. All data is presented when participants received more than one type of advice from the same source. For coding details, see table S7: numbers in parentheses correspond to the item number in that table.

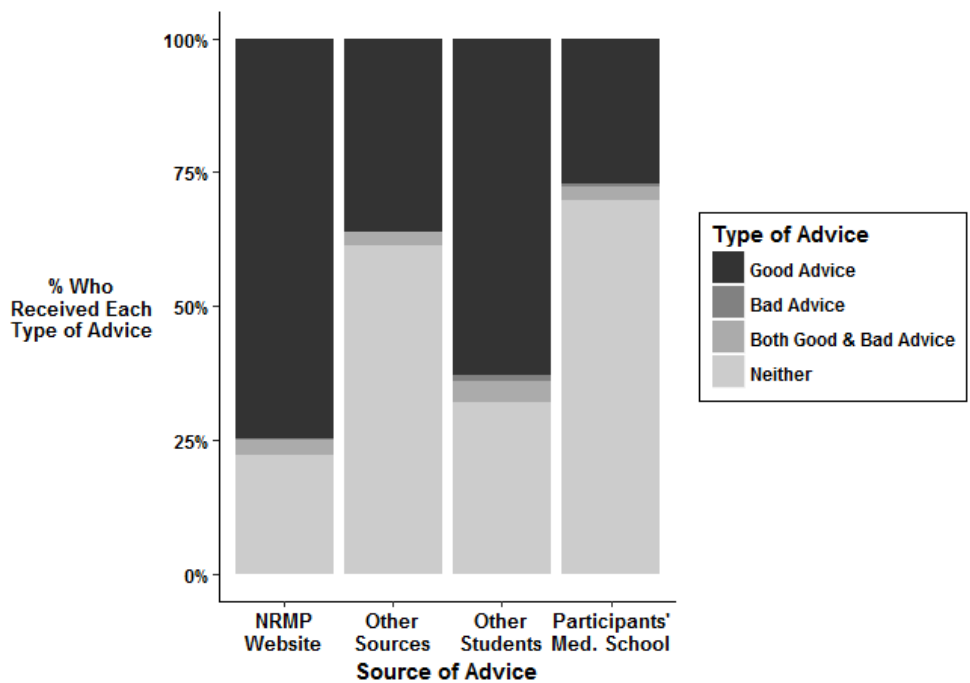

**Fig. S9. Providers of Good and Bad Advice.** Shown is the percent of good advice and bad advice participants report receiving from each source, conditional on receiving advice from that source. Six hundred ninety six participants (40.6% of sample) indicated receiving advice from the NRMP website, 404 participants (23.6% of sample) indicated receiving advice from other sources, 1,067 participants (62.3% of sample) indicated receiving advice from other medical students, and 1,227 (62.3% of sample) indicated receiving advice from their medical school.

    Good Advice: We classify participants as having received good advice from a source when they received advice that explained how the algorithm works, that the algorithm is student optimal, that they should rank order programs based on their true preferences, or that they should rank all programs at which they interviewed. (Items 1, 2, 3, and 4 in table S7).

    Bad Advice: We classify participants as having received bad advice from a source if they were advised to not rank "reach" programs, not rank programs they did not want to attend, not waste the top spots of their rank order list on "reach" programs, or over-rank programs that tell them they will be ranked highly (Items 5, 6, 7, and 8 in table S7).

    Both Good & Bad Advice: We classify participants as having received both good and bad advice if their free response was coded as containing at least one Good Advice item and one Bad Advice item.

    Neither: We classify participants as having received neither good nor bad advice if their free response was coded as containing none of the Good Advice or Bad Advice items.

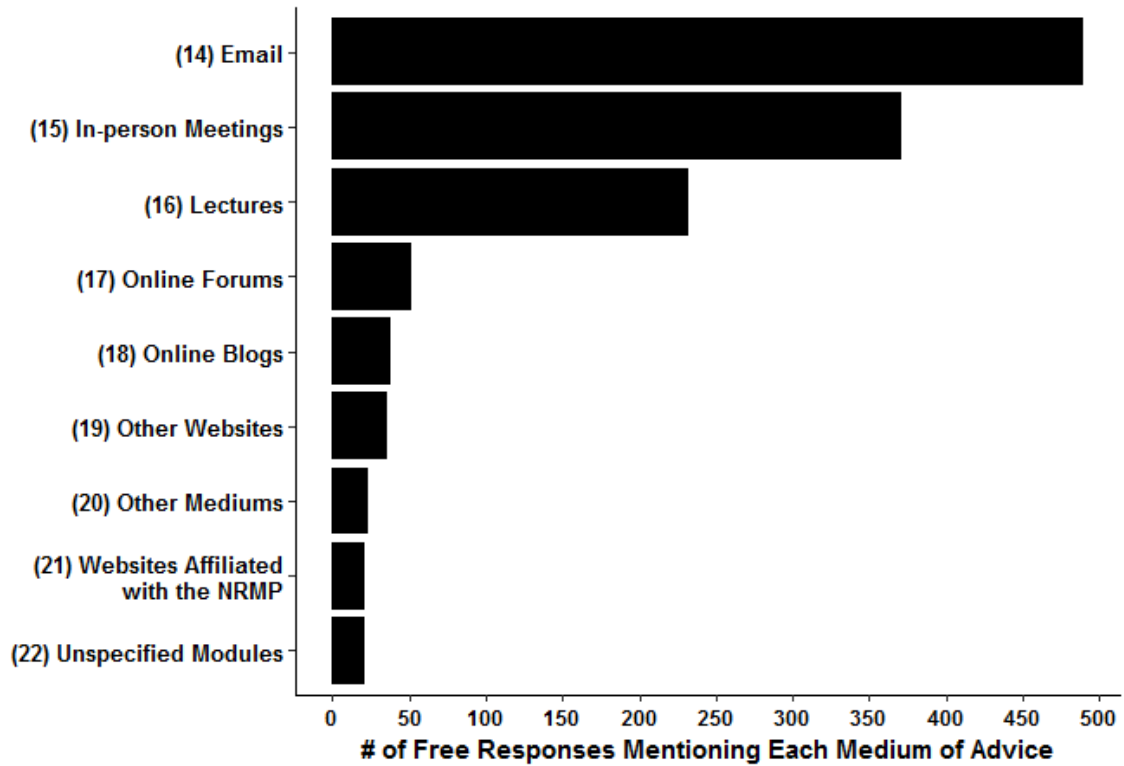

**Fig. S10. The Media That Sources Used to Provide Advice to NRMP Participants.** Participants' free responses were coded to detect nine different mediums through which participants received advice. This figure shows the frequencies that each medium was used to provide advice to participants. All mediums are counted when participants indicated that they received advice via multiple mediums from the same source. Free response prompts slightly varied by source. In particular, we explicitly asked participants to discuss how they received advice from their medical school. We did not attempt to elicit this information in the other three free response prompts. This explains, in part, why emails, in-person meetings and lectures are the modal mediums through which advice was provided. For coding details, see table S7: numbers in parentheses correspond to the item number in that table.

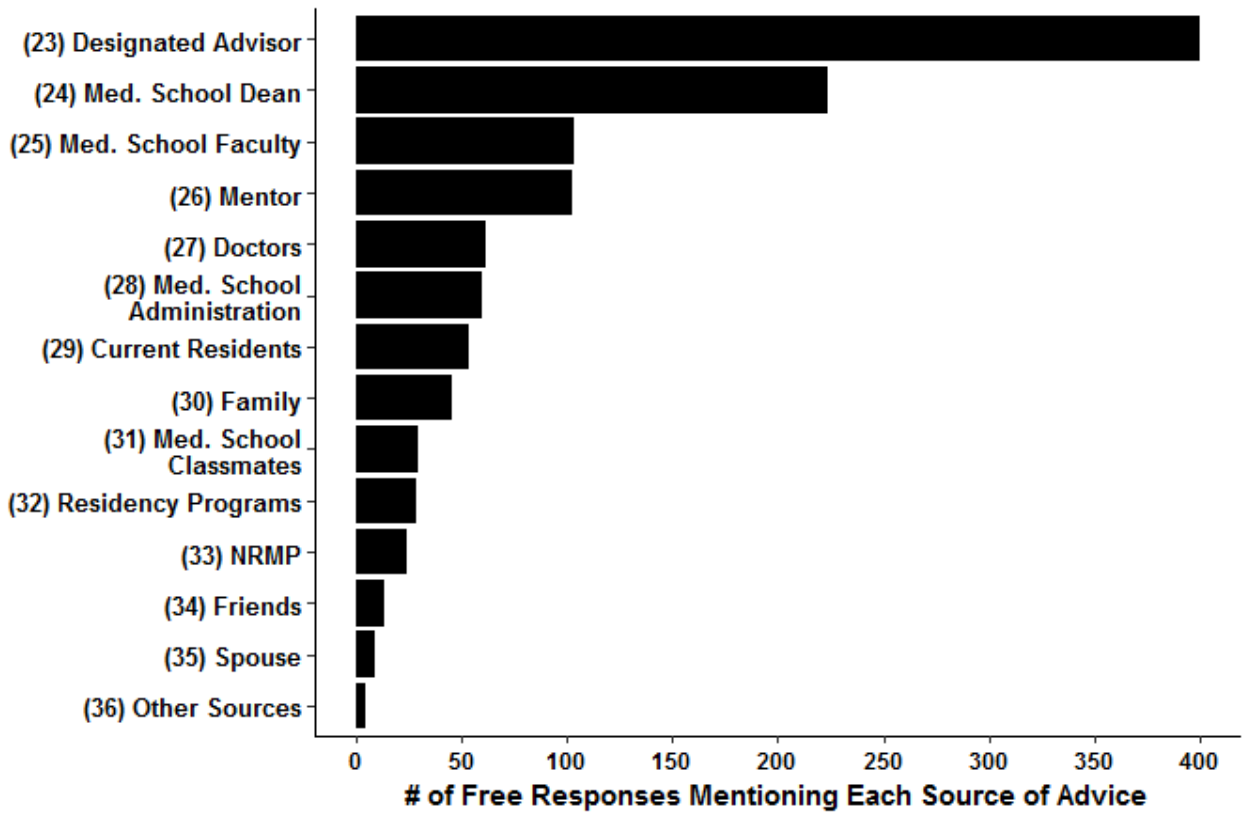

**Fig. S11. Providers of Advice.** Participants' free responses were coded to detect 14 different sources that provided advice to participants. This figure shows the number of free responses that mention each source. Only one source per participant-free-response observation is counted when participants indicated that they received advice via multiple sources in a single free response entry. Note that the free response prompts slightly varied by source. In particular, we explicitly asked participants to discuss who provided them with advice from their medical school and from other sources. We did not attempt to elicit this kind of information in the other two free response prompts. This explains, in part, why medical school affiliated sources are the modal sources of advice and why the NRMP is so infrequently mentioned. For coding details, see table S7: numbers in parentheses correspond to the item number in that table.

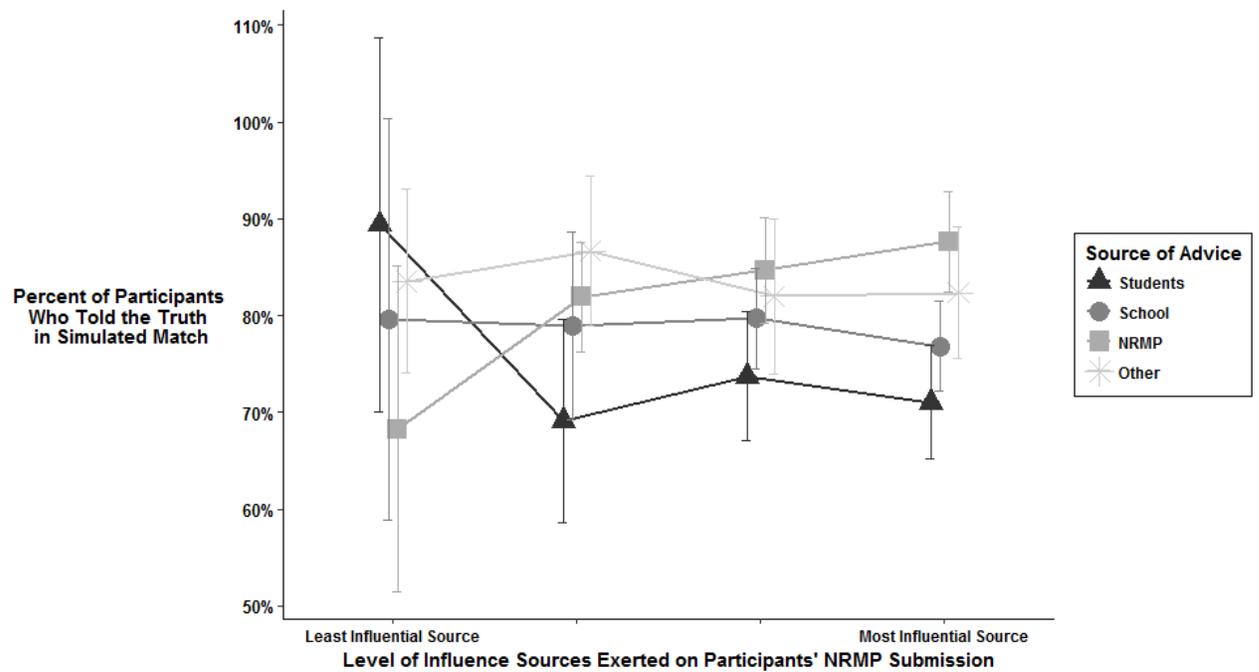

**Fig. S12. Source Influence.** Plotted are estimated percentages derived from a logit model predicting participants' likelihood to truthfully report preferences based on four categorical variables. These variables represent the level of influence each source of advice had on participants NRMP rank order list. After indicating which sources provided participants with advice, participants rank ordered each of these sources by the level of influence they exerted on their NRMP submission. Vertical lines at each data point show 95% confidence intervals.

**Table S1. Participating Medical Schools and Response Rates.** The Response Rate shows completed responses divided by number of graduates. The denominator is the number of eligible respondents (i.e., the number of students participating in the 2017 NRMP Match).

|  | State | University | Graduates in Class 2017 | Completed Responses | Response Rate |
|---|---|---|---|---|---|
| 1 | Indiana | Indiana University | 328 | 168 | 51.22% |
| 2 | Florida | University of South Florida | 162 | 140 | 86.42% |
| 3 | Texas | Texas Southwestern University | 227 | 139 | 61.23% |
| 4 | Massachusetts | Tufts University | 216 | 105 | 48.61% |
| 5 | Washington D.C. | The George Washington University | 181 | 101 | 55.80% |
| 6 | Ohio | The Ohio State University | 168 | 99 | 58.93% |
| 7 | Missouri | Saint Louis University | 181 | 98 | 54.14% |
| 8 | Wisconsin | University of Wisconsin | 171 | 79 | 46.20% |
| 9 | North Carolina | Wake Forest University | 116 | 79 | 68.10% |
| 10 | California | University of California, San Diego | 129 | 76 | 58.91% |
| 11 | Florida | University of Central Florida | 113 | 72 | 63.72% |
| 12 | Nebraska | University of Nebraska | 124 | 63 | 50.81% |
| 13 | Oklahoma | University of Oklahoma | 157 | 60 | 38.22% |
| 14 | New York | University of Rochester | 100 | 56 | 56.00% |
| 15 | New Jersey | Rowan University | 57 | 47 | 82.46% |
| 16 | Illinois | Southern Illinois University | 64 | 44 | 68.75% |
| 17 | South Carolina | University of South Carolina | 86 | 42 | 48.84% |
| 18 | Alabama | University of South Alabama | 70 | 41 | 58.57% |
| 19 | Michigan | Central Michigan University | 62 | 38 | 61.29% |
| 20 | Connecticut | University of Connecticut | 90 | 34 | 37.78% |
| 21 | Colorado | University of Colorado | 155 | 33 | 21.29% |
| 22 | South Dakota | University of South Dakota | 53 | 29 | 54.72% |
| 23 | Virginia | Virginia Polytechnic Institute and State University (Virginia Tech) | 41 | 26 | 63.41% |
| 24 | Georgia | Mercer University | 113 | 24 | 21.24% |
| 25 | Texas | Texas Tech University | 136 | 21 | 15.44% |
|  |  | **Total** | **3300** | **1714** | **51.94** |

**Table S2: Characteristics of Participating and Non-participating Medical Schools.**

We draw programs' information from the 2016 U.S. News & World Report database. This database contains demographic information for 93.9% (N=138) of all AAMC accredited medical schools in the United States and Puerto Rico, and all medical schools that participated in this study. Means are calculated after dropping schools with missing data. The "NR rate" shows the percentage of schools with missing data. The right-most column shows the *p* value of a two-sample t-test by participation status. Standard errors are in parentheses.

|  |  | All Programs | Non-Participating Programs | Participating Programs | P-value of Difference By Participation Status |
|---|---|---|---|---|---|
| Total Enrollment | Mean | 640.19 | 634.84 | 663.53 | 0.622 |
|  |  | (22.480) | (24.602) | (56.046) |  |
|  | NR rate | 26.09% | 26.55% | 24.00% | 0.795 |
|  |  | (0.038) | (0.042) | (0.087) |  |
| MCAT Composite | Mean | 32.09 | 32.15 | 31.84 | 0.644 |
|  |  | (0.255) | (0.304) | (0.353) |  |
|  | NR rate | 26.81% | 27.43% | 24.0% | 0.728 |
|  |  | (0.038) | (0.042) | (0.087) |  |
| Average Undergraduate GPA | Mean | 3.74 | 3.74 | 3.74 | 0.839 |
|  |  | (0.009) | (0.010) | (0.020) |  |
|  | NR rate | 26.09% | 26.55% | 24.00% | 0.795 |
|  |  | (0.038) | (0.042) | (0.087) |  |
| Acceptance Rate | Mean | 5.72 | 5.74 | 5.66 | 0.916 |
|  |  | (0.300) | (0.330) | (0.738) |  |
|  | NR rate | 26.09% | 26.55% | 24.00% | 0.795 |
|  |  | (0.038) | (0.042) | (0.087) |  |
| U.S. News Research Ranking | Mean | 43.55 | 41.86 | 50.59 | 0.198 |
|  |  | (2.667) | (3.067) | (4.956) |  |
|  | NR rate | 36.23% | 37.17% | 32.00% | 0.630 |
|  |  | (0.041) | (0.046) | (0.095) |  |
| Percent Female | Mean | 46.50% | 46.62% | 45.94% | 0.637 |
|  |  | (0.006) | (0.006) | (0.012) |  |
|  | NR rate | 4.35% | 4.42% | 4.00% | 0.926 |
|  |  | (0.017) | (0.019) | (0.040) |  |
| Observations |  | 138 | 113 | 25 |  |

**Table S3. Participant Demographics & Summary Statistics**

|  | Mean | SD |
|---|---|---|
| **Demographics and Respondent Characteristics** | | |
| Gender (female=1) | 0.48 | 0.500 |
| Age (years) | 27.20 | 3.628 |
| Participated in Couples Match (Yes=1) | 0.11 | 0.318 |
| Year MCAT was Taken | 2011.35 | 1.308 |
| Followed NRMP Training Link Provided in Experiment | 0.09 | 0.293 |
| **Summary Statistics** | | |
| # of Correctly Answered Raven's Matrices | 4.44 | 1.292 |
| Overconfidence in Raven's Performance | 0.69 | 0.462 |
| MCAT Score | 31.86 | 3.756 |
| Overconfidence in MCAT Performance | 0.12 | 0.330 |
| Received Advice from the NRMP Website | 0.41 | 0.491 |
| Received Advice from Medical Students | 0.62 | 0.485 |
| Received Advice from Medical School | 0.72 | 0.451 |
| Received Advice from Other Sources | 0.24 | 0.425 |
| Trust in Other Medical Students to Submit Truthful ROL | 0.37 | 0.482 |
| Trust in Residency Programs to Evaluate Students Fairly | 0.58 | 0.494 |
| Trust in the NRMP to Run the Algorithm Honestly | 0.97 | 0.162 |
| Observations | 1714 | |

**Table S4: Exclusions Prior to Data Analysis.** Presented are the observations excluded by each exclusion criteria across participating medical schools. Detailed descriptions of each criteria are listed in the Sample Exclusions subsection of Methods and Materials section of this document. Numbers in parentheses correspond to the column numbers. Column 9 shows the total number of excluded observations made at each medical school. Column 10 shows the data recovered after all exclusions.

| (1) Medical School | (2) Sample Collected | (3) Duplicates | (4) Nefarious Behavior | (5) Incomplete responses | (6) '17 NRMP Non-Participation | (7) Timing | (8) AWS | (9) Total Exclusions | (10) Analyzed Sample |
|---|---|---|---|---|---|---|---|---|---|
| Central Michigan | 46 | 0 | 0 | 6 | 0 | 0 | 2 | 8 | 38 |
| Colorado | 133 | 3 | 0 | 25 | 0 | 9 | 63 | 100 | 33 |
| GW | 129 | 0 | 0 | 21 | 4 | 3 | 0 | 28 | 101 |
| Indiana | 249 | 0 | 0 | 70 | 2 | 3 | 6 | 81 | 168 |
| Mercer | 72 | 0 | 0 | 27 | 1 | 2 | 18 | 48 | 24 |
| Nebraska | 76 | 0 | 0 | 8 | 1 | 1 | 3 | 13 | 63 |
| Ohio State | 124 | 0 | 0 | 22 | 0 | 2 | 1 | 25 | 99 |
| Oklahoma | 72 | 0 | 0 | 11 | 0 | 1 | 0 | 12 | 60 |
| Rochester | 68 | 0 | 0 | 8 | 0 | 0 | 4 | 12 | 56 |
| Rowan | 87 | 0 | 0 | 37 | 0 | 2 | 1 | 40 | 47 |
| Saint Louis | 188 | 8 | 24 | 53 | 0 | 3 | 2 | 90 | 98 |
| South Alabama | 49 | 0 | 0 | 7 | 0 | 1 | 0 | 8 | 41 |
| South Carolina | 49 | 0 | 0 | 4 | 2 | 1 | 0 | 7 | 42 |
| South Dakota | 32 | 0 | 0 | 3 | 0 | 0 | 0 | 3 | 29 |
| Southern Illinois | 54 | 2 | 0 | 7 | 0 | 1 | 0 | 10 | 44 |
| Texas Southwestern | 220 | 0 | 0 | 67 | 3 | 7 | 4 | 81 | 139 |
| Texas Tech | 45 | 0 | 0 | 3 | 0 | 1 | 20 | 24 | 21 |

Table S4 Continued

| | | | | | | | | | |
|---|---|---|---|---|---|---|---|---|---|
| Tufts | 131 | 1 | 0 | 22 | 1 | 1 | 1 | 26 | 105 |
| UCF | 95 | 0 | 0 | 21 | 0 | 1 | 1 | 23 | 72 |
| UConn | 92 | 0 | 0 | 26 | 0 | 5 | 27 | 48 | 34 |
| UCSD | 93 | 0 | 0 | 16 | 0 | 1 | 0 | 27 | 76 |
| USF | 280 | 4 | 0 | 123 | 3 | 10 | 0 | 140 | 140 |
| Virginia Tech | 43 | 2 | 0 | 8 | 0 | 1 | 6 | 17 | 26 |
| Wake Forest | 121 | 2 | 0 | 35 | 3 | 2 | 0 | 42 | 79 |
| Wisconsin | 89 | 0 | 0 | 10 | 0 | 0 | 0 | 10 | 79 |
| **Total** | **2637** | **22** | **24** | **640** | **20** | **58** | **159** | **923** | **1714** |

**Table S5. Suboptimal Play and Loss of Earnings.** Shown are the average dollar amounts participants earned for submitting optimal and suboptimal rank order lists in the simulated match. As in Figure 3, we show average earnings for each decile of participants' randomly assigned HST score. Column 6 shows the differences in mean earnings derived from a two-sample t-test. Standard errors are in parentheses. * $p < 0.05$, ** $p < 0.01$, *** $p < 0.001$.

| (1) Assigned HST Score Decile | (2) Sample Size | (3) Percent of Participants Who Submitted Optimal ROLs (%) | (4) Earnings From Suboptimal Behavior ($) | (5) Earnings From Optimal Behavior ($) | (6) Differences in Earnings ($) |
|---|---|---|---|---|---|
| 1 | 174 | 72.41 | 7.24 | 7.90 | -0.66*** (0.169) |
| 2 | 185 | 71.35 | 8.25 | 9.05 | -0.80*** (0.213) |
| 3 | 161 | 76.40 | 9.61 | 10.28 | -0.68 (0.356) |
| 4 | 167 | 71.86 | 11.97 | 12.13 | -0.16 (0.430) |
| 5 | 171 | 77.19 | 14.49 | 15.11 | -0.63 (0.621) |
| 6 | 171 | 75.44 | 18.99 | 18.72 | 0.27 (0.909) |
| 7 | 181 | 75.14 | 23.67 | 26.73 | -3.06* (1.456) |
| 8 | 168 | 80.95 | 29.06 | 32.54 | -3.47 (2.326) |
| 9 | 186 | 86.02 | 37.12 | 42.19 | -5.07* (2.508) |
| 10 | 150 | 80.67 | 42.24 | 48.35 | -6.11*** (1.571) |
| Total | 1714 | 76.72 | 18.20 | 22.80 | -4.60*** (0.847) |

**Table S6. Model with All Receipt-of-Advice Predictors.** Shown are the estimated average marginal effects derived from a logit model predicting truth-telling in the experimental task from the four measures of receiving advice. All predictors are binary. Standard errors in parentheses. * p < 0.05, ** p < 0.01, *** p < 0.001

|  | Average Marginal Effects (SE) |
|---|---|
| Received Advice from Medical Students | -0.01 |
|  | (0.023) |
| Received Advice from Medical School | 0.04 |
|  | (0.026) |
| Received Advice from the NRMP Website | 0.10*** |
|  | (0.021) |
| Received Advice from Other Sources | 0.07** |
|  | (0.023) |
| Observations | 1714 |

**Table S7: Free Responses Coding Items.** All free responses were coded on the 36 items presented in this table. All items were coded on a binary scale. Columns 1 and 2 list the item number and label which are referenced in figs. S8-S11. Column 3 defines each item. Column 4 lists the exact language that the research assistants were provided to code each free response. Column 5 shows an example of a free response that was coded as including the content of the item. See Free Response subsection in the Methods and Materials section of this document for more details on the coding procedure.

| (1) Item # | (2) Item Label | (3) Item Definition | (4) Coding Definition | (5) Example |
|---|---|---|---|---|
| 1. | Tell the truth | Participant received advice to tell the truth to the NRMP. | Tell the truth? (e.g., "Rank programs by preference") | "*Rank programs based on true preferences*" |
| 2. | Rank all programs | The participant received advice to rank all of the programs at which they interviewed. | Rank all programs (e.g., "rank every program", "apply broadly") | "*…List ALL programs where you would be willing to attend*" |
| 3. | The algorithm is student optimal | The participant was told that the algorithm used by the NRMP is applicant proposing. | Student optimal (e.g., "the system works in favor of students") | "*The algorithm is designed in the applicants favor...*" |
| 4. | How the algorithm works | The source provided information to the participant about how the algorithm works. | Informed how algo works | "*The website provided a thorough example and explanation of how the algorithm works*" |
| 5. | Don't rank reaches | The participant received advice to not rank the programs that they do not expect to be admitted to. | Don't rank places you cannot get into | "*…Rank places you most likely won't match at and put them in the 2$^{nd}$ and 3$^{rd}$ position and rank something you could possibly match at as number 1*" |
| 6. | Don't settle | The participant was advised to not rank programs that they would not want to attend. | Don't rank places you don't want to go | "*Rank only programs you'd be happy matching at...*" |
| 7. | Don't waste top picks | The participant was advised to rank programs they expected to match with at the top or their ROL. | Don't rank **highly** places you cannot get into | "*Don't rank too many 'reach' schools in the top 3*" |
| 8. | Rank programs that like you. | The participant was advised to rank programs that informed them that they were liked by the program. | Rank highly schools that tell me they like me | "*...rank certain places higher that said they ranked you highly.*" |
| 9. | Use external standards | The participant was advised to rank programs based on an external standard of quality. | Rank based on other external standards (e.g., "Droximity reports/rankings")[***] | "*To Rank programs based on their droximity rank as opposed to which program I liked the best.*" |

---

[***] Droximity is a website that ranks residency programs based on various program characteristics similar to the way US News & World Report ranks undergraduate universities.

|  |  |  |  |  |
|---|---|---|---|---|
| 10. | Advice on couples match | The participant received advice about the couples match. | Mention couples match | *"Online resources said that in the couples match is can be hard to navigate both partners' preferences"* |
| 11. | How to form preferences | The participant was advised on how to form their preferences. | Statement about how to form preferences | *"Consider future career plans, geographic location, competitiveness, how you meshed with the residents, and strengths of the program…"* |
| 12. | Not informative | The participant was provided platitudinous or nonsensical advice about the NRMP process. | non-informative (e.g., "go with gut", "do what you feel is right") | *"Follow your heart!"* |
| 13. | Other (advice) | The participant was given advice that did not fall into any of the above 12 items. | Other advice | *"I met with my advisors was given advice on strength of programs…"* |
| 14. | Email | The participant received an email about the NRMP. | Email | *"received emails and was recommended to read a blog written by medical school faculty"* |
| 15. | In-person meetings | The participant met with at least one individual, in person, to discuss the NRMP. | One-on-one meetings | *"I met with my advisory dean, along with 2 specialty advisors"* |
| 16. | Lectures | The participant attended a lecture that discussed the NRMP. | symposium (large meetings/lecture) | *"I attended a talk given by the dean regarding what types of programs to apply to…and how to apply to them"* |
| 17. | Online forums | The participant used online forums and message boards to learn about the NRMP. | Message-boards / forums | *"I used SDN forums to read about people's experiences with the match/interviewing in current and past years. …"* |
| 18. | Online blogs | The participant used online blogs or videos to learn about the NRMP. | Blogs, videos | *"I watched the video on the stable marriage algorithm in order to understand how the NRMP runs its algorithm…"* |
| 19. | Other websites | The participant used websites unaffiliated with the NRMP or the AAMC to learn about the NRMP. | Other websites | *"EMRA, AAFP, Doximity. These were all opinion pieces but provided concepts for considering rank list."*[†††] |
| 20. | Other mediums | The participant used other mediums not specified in items 14-19, 21, or 22 to gather information about the NRMP. | other mediums | *"Doximetry newsletter. They said that it is mostly about geography and going with your gut…."* |
| 21. | Unspecified modules | The participant used online learning modules to learn about the NRMP. | Unspecified modules | *"A self-learning module was provided to us explaining the match algorithm and how to rank programs."* |

---

[†††] The Emergency Medical Residency Association (EMRA) and The American Academy of Family Physicians (AAFP) are websites designed to provide medical students, residents, and physicians information about emergency medicine and family medicine, respectively.

| | | | | |
|---|---|---|---|---|
| 22. | Websites affiliated with the NRMP | The participant gathered information about the NRMP from AAMC or NRMP affiliated websites. | Affiliated Websites (e.g., NRMP.org, AAMC.org) | *"AAMC, AAFP. Advice was to rank according to true preference."* |
| 23. | Designated advisor | The participant was assigned an advisor from their medical school to navigate the NRMP process. | Designated advisor | *"I was assigned an advisor for the residency application process, plus I had a separate advisor whom I trusted from my rotations ..."* |
| 24. | Med. school dean | The participant received advice from their medical school dean about the NRMP. | Dean | *"I received emails, plus we had meetings with the dean of students. He basically just answered any questions we had and told us all of the standard advice."* |
| 25. | Med. school faculty | The participant received advice from a faculty member affiliated with their medical school about the NRMP. | Faculty | *"I received advice from faculty that I trust who are also in the specialty that I applied to in the match."* |
| 26. | Mentor | The participant received advice from a mentor (from their medical school or otherwise) about the NRMP. | Mentor | *"I received advice from my surgery mentor"* |
| 27. | Doctors | The participant received advice from practicing physicians about the NRMP. | practicing doctors | *"Community physicians recommended ranking most highly those places I believes I had the best chance of matching"* |
| 28. | Med. school administration | The participant was contacted by or sought information from their medical school administration about the NRMP. | School administration (e.g. "the school sent me an email") | *"Emails from administration that detailed statistics on what ultimately made residents happy with their decisions."* |
| 29. | Current residents | The participant received advice from current residents about the NRMP. | current residents | *"current residents: they told me to tell my #1 choice that I was ranking them"* |
| 30. | Family | The participant received advice from members of their family about the NRMP. | Family | *"My parents gave me advice and they told me to rank based on my goals and where I will be happy."* |
| 31. | Med. school classmates | The participant received advice from their fellow medical school classmates about the NRMP. | classmates | *"Fellow students in my class. They repeated the same advice as the NRMP: to rank programs in their true order."* |
| 32. | Residency programs | The participant received advice from one or more residency programs about the NRMP. | Residency programs | *"On many interviews, the PDs told us to rank every program we interviewed at and to rank them based upon where we feel we would fit in best..."* |
| 33. | NRMP | The participant received advice from the NRMP (either directly through their website or via school | NRMP | *"I received countless emails, often citing NRMP sources,"* |

|     |               | administration emails) about the NRMP. |               |                                                                                                      |
| --- | ------------- | -------------------------------------- | ------------- | ---------------------------------------------------------------------------------------------------- |
| 34. | Friends       | The participant received advice from their friends about the NRMP. | Friends | *"Friends. Basically telling me to do whats important for me."* |
| 35. | Spouse        | The participant received advice from their spouse about the NRMP. | Spouse | *"My wife--she advised me to stay here, as she is a PhD student and still has a few years left in er program…"* |
| 36. | Other sources | The participant received advice from sources not specified in items 23-35. | Other sources | *"Original peer-reviewed article describing match's mathematical algorithm"* |

**Table S8. Model with All Trust Predictors.** Shown are the estimated average marginal effects derived from a logit model predicting truth-telling in the experimental task using the three measures of trust in other market participants. All predictors are binary. Standard errors in parentheses. * $p < 0.05$, ** $p < 0.01$, *** $p < 0.001$

|  | Average Marginal Effects (SE) |
|---|---|
| Trust in Residency Programs to Evaluate Students Fairly | -0.05* (0.021) |
| Trust in Other Medical Students to Submit Truthful ROLs | 0.01 (0.022) |
| Trust in the NRMP to Run the Algorithm Honestly | 0.06 (0.069) |
| Observations | 1714 |

**Table S9. Predictors of Truth-Telling Status.** Shown are the logit estimates plotted in Figure 4, and a replication of these estimates using OLS. Models 1 and 3 present estimates obtained from the complete model, including the entire battery of predictors. Models 2 and 4 presents the estimate for each univariate model, predicting truth-telling with only the single variable represented in that row. Standard errors are in parentheses. *$p < 0.05$, ** $p < 0.01$, *** $p < 0.001$.

|  | AME Derived from Logit Model | | Beta Coefficients Derived from OLS | |
| --- | --- | --- | --- | --- |
|  | (1) | (2) | (3) | (4) |
| Model: | Complete | Univariate | Complete | Univariate |
| Strategic Position in Market | 0.03** | 0.04*** | 0.03** | 0.04*** |
|  | (0.010) | (0.010) | (0.010) | (0.010) |
| Cognitive Ability | 0.05*** | 0.03** | 0.05*** | 0.03** |
|  | (0.012) | (0.010) | (0.012) | (0.010) |
| Overconfidence | 0.08** | 0.02 | 0.08** | 0.02 |
|  | (0.028) | (0.022) | (0.027) | (0.022) |
| Salient Expectations | -0.01 | -0.01 | -0.01 | -0.01 |
|  | (0.020) | (0.020) | (0.020) | (0.020) |
| Medical School Advice | 0.04 | 0.08*** | 0.04 | 0.08*** |
|  | (0.025) | (0.024) | (0.025) | (0.023) |
| NRMP Advice | 0.10*** | 0.12*** | 0.09*** | 0.12*** |
|  | (0.021) | (0.020) | (0.022) | (0.021) |
| Student Advice | -0.01 | 0.04* | -0.01 | 0.04* |
|  | (0.023) | (0.021) | (0.023) | (0.021) |
| Other Advice | 0.07** | 0.09*** | 0.07** | 0.09*** |
|  | (0.023) | (0.022) | (0.024) | (0.024) |
| Residency Trust | -0.05* | -0.05* | -0.05* | -0.05* |
|  | (0.021) | (0.020) | (0.021) | (0.021) |
| NRMP Trust | 0.05 | 0.05 | 0.05 | 0.05 |
|  | (0.066) | (0.067) | (0.062) | (0.063) |
| Student Trust | 0.02 | 0.00 | 0.02 | 0.00 |
|  | (0.021) | (0.021) | (0.022) | (0.021) |
| Observations | 1714 | 1714 | 1714 | 1714 |